\documentstyle{article}

\def\avg#1{\langle {#1} \rangle }
\def\half{\mbox{$\frac{1}{2}$}}
\def\third{\mbox{$\frac{1}{3}$}}
\def\F#1#2{F_{#1#2}}
\def\Uplaq#1#2{{U_{#1#2}^{\rm plaq}}}
\def\Urect#1#2{{U_{#1#2}^{1\times2}}}
\def\U#1{U_{#1}}
\def\Ut#1{\tilde{U}_{#1}}
\def\Udag#1{{U^\dagger_{#1}}}
\def\A#1{A_{#1}}
\def\Tr{{\rm\,Tr\,}}

\newenvironment{exercise}{\small\begin{description}
                         \item[{Exercise:}]}{\end{description}}

\newenvironment{refer}{\small\begin{quote}}{\end{quote}}

\newcommand{\be}{\begin{equation}}
\newcommand{\ee}{\end{equation}}
\newcommand{\eq}[1]{Eq.~(\ref{#1})}
\newcommand{\e}{{\rm e}}

\newcommand{\nl}{\nonumber \\}
\newcommand{\order}{{\cal O}}
\newcommand{\lag}{{\cal L}}
\newcommand{\lat}{{\rm lat}}

\newcommand{\xv}{{\bf x}}
\newcommand{\rv}{{\bf r}}
\newcommand{\xhat}{{\bf \hat{x}}}
\newcommand{\fm}{{\rm fm}}
\newcommand{\mev}{{\rm MeV}}
\newcommand{\su}{{${\rm SU}_3$}}
\newcommand{\msb}{{\rm\overline{MS}}}

\newcommand{\Bv}{{\bf B}}
\newcommand{\pv}{{\bf p}}
\newcommand{\Ev}{{\bf E}}

\newcommand{\delv}{{\bf \del}}
\newcommand{\sigmav}{{\mbox{\boldmath$\sigma$}}}
\newcommand{\delfour}{{\Delta^{(4)}}}
\newcommand{\Mbz}{{M_b^0}}
\newcommand{\del}{{\bf \Delta}}
\newcommand{\delsq}{\Delta^{(2)}}

\title{Lattice QCD for Small Computers\thanks{
Lectures presented at TASI~93 (Boulder, June 1993), and at the
UK Summer Institute (St.~Andrews, Aug 1993).}}
\author{G.Peter Lepage \\
\small Floyd R. Newman Laboratory of Nuclear Studies \\
       \small Cornell University, Ithaca, NY 14853 \\
		    {\small gpl@hepth.cornell.edu} }
\date{\small February 1994 (revised)}

\begin{document}

\maketitle

\section{Introduction}
Despite recent trends towards ever larger computers, I believe that
accurate and reliable lattice simulations of QCD are possible on quite
modest computers, possibly even on the workstation or PC sitting on your
desk. If true, this will have a profound effect on the way we deal with
strong interactions, both experimentally and theoretically. In these
lectures I outline the reasons behind my belief.

The central issue is the reliability of lattice simulations of QCD. Lattice
simulations, like experiments, have two sources of error: statistical errors
due to our use of Monte Carlo methods for doing the simulations; and
systematic errors due to the various approximations we make to simplify the
theory to the point that it is numerically tractable. Statistical errors are
pretty much understood, and under control; I do not discuss them further in
these lectures. Instead, I concentrate on systematic errors. It is the
control of systematic errors that determines the scale and reliability of
contemporary QCD simulations.

I begin in Section~2 with a discussion of the main sources of systematic
error in lattice simulations. Of these errors those due to the finite lattice
spacing are the most important. I describe a general strategy for
systematically improving lattice lagrangians to remove
lattice-spacing errors. This is an alternative to simply reducing the lattice
spacing that promises to be much more efficient. This strategy relies upon
weak-coupling perturbation theory, and so in Section~3 I discuss the extent
to which perturbation theory can be trusted in lattice QCD. Contrary to much
conventional lore lattice perturbation theory seems to work well even at
distances as large as~1/2~fm. In Section~4 I discuss another obstacle to the
perturbative improvement of lattice QCD. The problem is with ``tadpole''
terms that arise in perturbation theory when computing corrections to the QCD
lagrangian. The tadpoles are usually too large to be dealt with
perturbatively. All such contributions have a common origin, and I describe
a simple nonperturbative procedure for dealing with them.  In Section~5 I
illustrate all of these techniques with recent simulation results for
the $\Upsilon$~family of mesons\,---\,an area where lattice~QCD has
already largely succeeded. Finally, in Section~6, I summarize the central
result of these lectures.

\section{Systematic Errors in Lattice QCD}
There are three important sources of systematic error in lattice QCD:
\begin{enumerate}
\item[a)] finite volume errors that arise because simulations cover only a
finite volume~$V=L^4$ in space-time;
\item[b)] errors due to extrapolations from large quark masses, where the
simulations are easy, down to realistic quark masses for $u$~and $d$~quarks,
or, equivalently, from large $m_\pi^2\propto m_{u,d}$ to $m_\pi^2  = ({\rm
140~MeV})^2$;
\item[c)] errors due to the finite lattice spacing~$a$ in the grid used to
approximate continuous space-time.
\end{enumerate}
The central question in numerical QCD then is: How large must $L$~be, and how
small $m_\pi$ and $a$, for simulations to be realistic (and useful)? The
answer to this question determines the cost and feasibility of lattice
simulations.

The cost of a lattice simulation is at least proportional to the number of
sites in the grid,
\be
\mbox{number of sites} = \left( \frac{L}{a}\right)^4.
\ee
Usually, however, it is more costly because of a phenomenon called
``critical slowing down'' that tends to slow the convergence of the
iterative algorithms used in lattice simulations. A more realistic estimate
of the cost is
\be \label{cost}
\mbox{cost} \propto \left(\frac{L}{a}\right)^4
  \left(\frac{1}{a^2}\right)^\alpha \left(\frac{1}{m_\pi^2 a}\right)^\beta
\ee
where typically $\alpha$ and $\beta$ are somewhere in the range 0--1
(usually closer to 1 than 0).  Obviously this cost is highly sensitive to the
values of $a$, $m_\pi$, and $L$; for example, in the worst case, the cost is
proportional to the seventh power of $1/a$. Thus it is critically important
to make $a$ and $m_\pi$ as large as possible, and $L$ as small as possible.
We cannot afford to be sloppy or to ``play it safe.''

Here I discuss each of these systematic errors in turn. However my emphasis is
on the finite-$a$ errors as these are by far the most crucial determinants of
the cost (see \eq{cost}), and they are also the most misunderstood of the
systematic errors.

\subsection{Finite $L$}
Clearly when we simulate a hadron, like the pion or the proton, that is
roughly 2~\fm\ in diameter, we must have $L$ at least as large as 2~fm.The
nature of the systematic error when $L\ge2$~\fm\ is easily understood.
Lattice simulations generally use periodic boundary conditions to define the
differential operators in the  action. This means that one is never studying
just a single hadron on a grid, but rather an infinite crystal of hadrons
consisting of the hadron and images of it, induced by the boundary
conditions, at intervals of $L$ in each direction. The properties of our
hadron are affected by interactions with all of its images. Its energy,
for example, is shifted by the interaction energy between the hadron and its
images. The interaction energy between two hadrons a distance~$r$ apart
is
\be
K\,\frac{\e^{-m_\pi r}}{r}.
\ee
Thus the finite volume error falls off  exponentially quickly with
increasing~$L$ provided $L$~is large enough. This will not be the case when
$L$~is so small that the hadron overlaps with its image; then one finds
powerlaw effects related to the geometrical volume of the overlap. But for
$L\ge2~\fm$ we expect
\be
\mbox{finite $L$ errors} \propto \e^{-m_\pi L} \approx \e^{-L/1\:\fm}
\ee

This formula suggests that finite volume errors for hadrons ought to be
10\% or less for $L$'s of order 2~\fm\ or larger, and this seems to be
the case in current simulations.
Even if 2~fm is too small, the exponentially-falling errors vanish so
quickly with increasing~$L$ that it is almost certain that 2.5~or 3~fm will
suffice. The combination of exponentially small errors and powerlaw cost
means that finite-volume effects are both tractable and fairly well
understood.

\subsection{Large $m_\pi$}
The numerical intractability of small quark masses means that simulations
rarely have pion masses smaller than about 500~\mev. (Remember that $m_\pi^2
\propto m_{u,d}$ is a measure of the quark mass.) As a consequence, it is
standard practice in lattice simulations to present results as functions of
$m_\pi$, extrapolating down to the physical mass~(140~\mev). The
reliability of this procedure depends upon how sensitive the simulation
results are to the pion mass.

We have, in chiral perturbation theory, a reasonably reliable tool for
exploring the $m_\pi$ dependence of different quantities. With chiral
perturbation theory we can compute the $m_\pi$~dependence of various
quantities. Some, like the nucleon mass, are quite sensitive to~$m_\pi$
when $m_\pi\approx 500$~MeV. Others, like  the pion decay constant~$f_\pi$,
are relatively insensitive.
This suggests that extrapolations in $m_\pi$ ought to work well for $f_\pi$,
while extrapolations of the nucleon mass could lead to problems.

Chiral perturbation theory can help us decide which quantities are least
sensitive to $m_\pi$, and we might wish to concentrate our attention on
these. More importantly, we should push on to smaller
$m_\pi$'s. Large $m_\pi$ errors are typically be of order
\be
\mbox{large $m_\pi$ errors} \approx \left(\frac{m_\pi}{700~\mev}\right)^2 .
\ee
This suggests that $m_\pi$'s of order 250~MeV or less ought to result in
systematic errors that are less than 10\%. Given that the cost grows only as
$1/m_\pi^2$, it should be possible to work at these lower masses,
especially if we are able to use significantly larger lattice spacings.

\subsection{Finite $a$\,---\,Classical Field Theory}
In thinking about errors due to the finite lattice spacing, I find it useful
to look first at a classical field theory on a lattice. Consider, for
example, Poisson's equation:
\be
\partial^2\phi(\rv) = \rho(\rv).
\ee
On a grid, we replace derivatives by differences, for example
\be
\partial^2_x\phi(\rv) \to
\frac{\phi(\rv+a\xhat)-2\phi(\rv)+\phi(\rv-a\xhat)}{a^2},
\ee
and specify $\phi$ only at the nodes of a lattice with spacing~$a$.

Obviously $a$ must be small compared with the size of any important
structure in~$\rho$ or~$\phi$. If this is the case, the finite-$a$ errors
fall off like a power of~$a$. This is easily demonstrated. The difference
operator that replaces the second derivative is defined by
\begin{eqnarray}
a^2\Delta^{(2)}_x \phi(\rv) &\equiv&
   \phi(\rv+a\xhat)-2\phi(\rv)+\phi(\rv-a\xhat) \nl
 &=& \left(\e^{a\partial_x} -2 + \e^{-a\partial_x}\right)\,\phi(\rv).
\end{eqnarray}
Expanding the exponentials, we find that
\be
\Delta_x^{(2)} \phi(\rv)
\approx \left(\partial^2_x + \frac{a^2\partial_x^4}{12} +\cdots\right)\,
 \phi(\rv).
\ee
Thus by replacing $\partial^2$ with $\Delta^{(2)}$ we  make
errors of order
\be
\mbox{finite $a$ errors} \approx \frac{a^2\overline{p}^2}{12}
\ee
where $\overline{p}$ is the typical wavenumber (ie, momentum) in the Fourier
transform of~$\phi(x)$.

There are two lessons to learn from this example. The first is that only a
small number of lattice points is needed to obtain precision of order 10\%.
For example, if $\phi$ consists of a single smooth bump of width~$w$, then a
lattice spacing $a\approx w/3$ gives roughly 10\%~accuracy:
$\overline{p}\approx2\pi/\lambda$, where $\lambda\approx 2w \approx 6a$,
implies that $a^2\overline{p}^2/12 \approx 0.08$.
A general rule of thumb is that it takes roughly $3^d$~grid points per
$d$-dimensional bump to begin to model that bump accurately on a grid.

The second lesson is that one should not reduce the lattice spacing much
below $w/3$ when trying to improve the accuracy of the grid approximation.
This is because reducing the lattice spacing greatly increases the cost of
the simulation while only  modestly improving the errors. For
example, to reduce 10\%~errors to 1\%~errors using our
discretization of the three-dimensional Poisson's equation one would have to
decrease $a$ by a factor of 3, thereby increasing the number of grid points
by a factor of $3^3 = 27$. Taking account of critical slowing down, the cost
of the simulation would increase by a factor of 80--200.

A far more efficient way to reduce errors is to improve the discretization,
keeping the lattice spacing large. For example, from our discussion
above,
\begin{eqnarray}
\partial^2_x &=&
   \Delta^{(2)}_x-\frac{a^2\partial^4_x}{12}-\frac{a^4\partial^6_x}{360}
    - \cdots \nl
&=& \Delta^{(2)}_x -\frac{a^2(\Delta^{(2)}_x)^2}{12} +
  \frac{a^4\partial^6_x}{90} +\cdots \label{clcorr}
\end{eqnarray}
and this implies that the lattice equation
\be
\sum_i \left(\Delta^{(2)}_i -
\frac{a^2(\Delta^{(2)}_i)^2}{12}\right)\,\phi(\xv)
 = \rho(\xv)
\ee
is accurate up to corrections of order~$(a\overline{p})^4/90$.
This improved lattice equation gives roughly 1\%~accuracy with $a=w/3$.
The added complexity of the equation increases the cost of simulating, but only
by a factor of~1.5 or~2.

Thus for classical fields we conclude that coarse lattices are
optimal\,---\,perhaps $3^d$ grid points per bump in the answer. High precision
is obtained by improving the lattice operators from which the field equations
are constructed,  not by reducing the lattice spacing.

\begin{exercise}
Compare exact derivatives with lattice derivatives for $\phi(x) =
\exp(-x^2/2)$ in one dimension. Try lattice spacings of~$a=0.5$ and~$1$, and
study the second derivative at $x=0$. Compare the errors obtained using the
leading-order lattice derivative and the improved derivative. Check how
these errors scale with~$a$.

The lattice derivatives can be corrected to all orders in~$a$ by continuing
the process outlined above. Compute the error in the
infinite-order lattice approximation of $\partial_x^2\phi(0)$. In general,
for what sorts of function~$\phi(x)$ will the error be exponentially
small, power-law suppressed, or identically zero? (Hint: Use Fourier
transforms. Fourier transforming the lattice operators is
simple. In particular, the infinite-order lattice approximation to
$\partial_x^2$ has transform $-k^2$, where $k$~is the wavenumber.)

This last exercise shows that our lattice derivatives can never be perfect;
typically there are residual non-powerlaw differences between
them and continuum derivatives. Examine the possibility of removing
even these errors when~$r\gg a$ in our lattice Poisson's equation with
a source~$\rho(\rv)$ that is localized
around~$r=0$. In particular, consider adding local counterterms to the
source: $\rho(\rv) \to \rho + c_0\,\delta(\rv) +
c_2\,\partial^2\delta(\rv) + \cdots$. (Note the analogy with multipole
expansions.)
\end{exercise}

\begin{exercise}
Compute the ground-state energy of a nonrelativistic particle trapped in a
box of length $L$ using the Shr\"odinger equation with discretized
derivatives of the sort discussed above. (This is easily done analytically
using Fourier series for the eigenfunctions.)  Verify the ``3 points per
bump'' rule and explore the utility of improving the discretization as we
did above.

As a variation, consider adding a harmonic potential to the hamiltonian
and solving the eigenvalue problem numerically. Defined on a grid, the
hamiltonian operator becomes a matrix, and the energy eigenvalue problem a
simple matter of determining the eigenvalues of that matrix. Again explore
the relative merits of reducing the lattice spacing versus improving the
discretization. Try the problem in 2 or more dimensions as well.
\end{exercise}

\subsection{Finite $a$\,---\,Quantum Field Theory}
The analysis of finite-$a$ errors for quantum theories is complicated by
the fact that quantum fields are rough at all length scales. This roughness
is due to quantum fluctuations. In perturbation theory, it is responsible
for the ultraviolet sensitivity and infinities of loop diagrams. More
generally, it seems to call into question the validity of a discrete
approximation. Indeed it is hard to imagine defining
continuum derivatives for an infinitely rough field, let alone discrete
approximations to them.

Somewhat surprisingly, finite-$a$ errors in quantum theories are rather
similar to those in classical theories. This is because
only the long-wavelength structure is physical on a
lattice. In general, any long-wavelength, low-momentum probe is sensitive
only to fields averaged over a region of order the size of the probe.
This averaging smoothes out quantum fluctuations.
Consequently the long-wavelength or infrared behavior of
a quantum theory is insensitive to the details of its
short-wavelength or ultraviolet behavior; and thus there are infinitely many
theories, each with different ultraviolet dynamics, that give identical or
nearly identical infrared physics. We take advantage of this fact when we
construct our lattice theory: we replace the correct ultraviolet
dynamics of the continuum theory with the incorrect, but numerically
tractable, dynamics of the lattice. By changing its bare coupling constants,
we are able to adjust the dynamics of the lattice theory so that its
infrared behavior is the same as in continuum up to corrections that
vanish with the lattice spacing~$a$\,---\,just as in classical lattice
theories.

To see how this works consider a $\phi^4$~field theory defined on a lattice:
\be
\lag_\lat = -\frac{1}{2} \,\sum_\mu \phi \Delta^{(2)}_\mu \,\phi +
    \frac{1}{2}\, m^2\phi^2 + \frac{1}{4!}\,\lambda \phi^4
\ee
This lagrangian, used in a path integral, defines a euclidean
lattice approximation to the continuum quantum theory.
Classically, large distortions due to the grid are restricted to the
ultraviolet modes; but, in the quantum theory, the infrared
sector is strongly affected as well since it couples to the ultraviolet
modes through quantum fluctuations (ie, loops in perturbation theory). For
example, distortion of the ultraviolet modes changes the mass
renormalization, thereby shifting the renormalized mass of the
$\phi$~particle\,---\,a significant modification of the infrared behavior of
the theory.

Introducing a grid causes $\order(1)$~changes in the behavior of the
quantum theory at all length scales, but the continuum behavior at large
distances can always be restored by shifting or ``renormalizing'' the bare
coupling constants. The renormalized mass of the $\phi$~particle, for
example, is easily restored to its proper value by adjusting the bare
mass~$m$ in the lattice action.
The grid is nothing more than an ultraviolet cutoff, restricting the path
integral to momenta smaller than~$\pi/a$. Renormalization theory tells us that
the errors caused in, say, a scattering amplitude by introducing a finite
cutoff~$\Lambda$ are suppressed by powers of~$p_{\rm ext}/\Lambda$,
where $p_{\rm ext}$ is the largest {\em external\/} momentum in the
amplitude. This is true provided the coupling constants of the theory are
shifted appropriately from their continuum values. Thus, for our lattice
$\phi^4$~theory, there are particular values of~$m$ and~$\lambda$ such
that the lattice result for any physical quantity is the same as in the
continuum theory up to corrections that vanish as powers of~$a\overline{p}$,
where $\overline{p}$ is a momentum associated with the classical (as opposed
to quantum) size and scales relevant to that quantity. Generally the values
needed for~$\lambda$ and~$m$ depend upon the lattice spacing~$a$ (the
couplings are said to ``run'' with varying~$a$).

The problem of dealing with the effects of complex short-distance structure
on long-distance behavior arises in many contexts, both quantum and
classical. A classical example is the formulation of electrostatics inside a
dielectric medium (a piece of glass, for example). The microscopic electric
field inside a dielectric is very complicated and fluctuates rapidly even
over distances as short as 1~\AA. Usually, however, it is not the
microscopic field that is physically relevant. Any macroscopic probe of the
electric field sees only the average  field, averaged over a large region
that includes many atoms. This averaging smoothes out the rapid
fluctuations,  producing a relatively smooth macroscopic field that is
described by simple equations. The macroscopic equations  are universal in
form; the type and nature of the atoms making up the dielectric are largely
irrelevant. The only information needed about the microscopic structure of
the material is contained in the numerical value of its dielectric constant.
The dielectric constant is analogous to the bare couplings in our lattice
theory; and the standard formalism for dielectrics comes from a
straightforward application of renormalization ideas to the classical
problem.

Despite quantum fluctuations, the finite-$a$ errors in our quantum theory are
quite similar to those in the classical theory. Also as in the classical
case, we can correct the theory to systematically reduce these errors.
Renormalization theory tells us that all errors of order~$(a\overline{p})^n$
can be removed by adding local interactions of dimension~$4+n$ and lower to the
lattice lagrangian with appropriate couplings (and readjusting any couplings
already present in the lagrangian). The correction terms are local because they
are correcting for the physics at distances shorter than the lattice spacing.
Only terms that preserve the symmetries of the theory need be added, and so
there aren't very many for small~$n$. Our lattice $\phi^4$~theory, for example,
has no possible correction term of dimension five. This means that the
leading finite-$a$ errors are quadratic in~$a$.
There are only two dimension six correction terms, and so the
$\order(a^2\overline{p}^2)$~errors in $\lag_\lat$ are removed by adding
\be
\delta\lag_\lat \equiv \frac{a^2 c(a)}{24}\,\sum_\mu
    \phi (\Delta^{(2)}_\mu)^2 \phi +
  \frac{a^2 d(a)}{6!}\,\phi^6
\ee
and by further shifting~$m(a)$ and~$\lambda(a)$.

\begin{exercise}
There are actually two other dimension-6 operators that are consistent with
the symmetries of the  lattice $\phi^4$~theory:
$\left(\Delta^{(2)}\phi\right)^2$ and~$\phi^3\Delta^{(2)}\phi$. We do not
include these in our corrected lagrangian because they are redundant. These
operators can always be removed from the lagrangian by making the replacement
\be
\phi(x) \to \phi(x) \equiv \tilde\phi(x) +a^2 f_1\, \phi^3(x)
+a^2 f_2\,\Delta^{(2)}\phi(x)
\ee
in the lagrangian, where~$f_1$ and~$f_2$ are suitably chosen functions of
the bare mass and couplings. The path integral for the lattice theory is an
ordinary multidimensional integral, and this field redefinition is merely a
change of integration variable; it cannot alter the physical content of the
theory.

Show that~$f_1$ and~$f_2$ can always be adjusted so that both redundant
operators cancel when~$\phi$ is transformed in the lagrangian. This is easy
to see at tree-level (ie, in the classical theory). What complications
arise in one-loop order? For example, what happens to the jacobian
associated with the field transformation?

While a field transformation of this sort has no effect on physical
quantities, it usually changes off-shell Green's functions. Thus the
Green's functions of our improved theory may still deviate from the continuum
Green's functions in order~$a^2$. However particle masses, on-shell
scattering amplitudes, and the like will all be accurate through
order~$a^4$. This is all that we need. Examine the differences between
on-shell and off-shell quantities using tree-level perturbation theory for
equivalent theories with and without the redundant operators.

An important step in discretizing a quantum field theory for
nonperturbative analyses is to verify that the lattice version of the
theory is stable. This is an issue when improving the lagrangian.
For example, if it happened that the $\phi^6$~coupling, $d(a)$, was negative,
we would have to worry about stabilizing the improved theory against
$\phi\to -\infty$. Luckily we have tremendous latitude in the design of
our corrections. For example, we can freely  add $\phi^8$,
$\phi^{10}$\ldots{}terms to the lagrangian without affecting simulation
results through order~$(a\overline p)^4$ (provided all couplings are
readjusted appropriately). Such terms could be used to stabilize the
discrete theory. The presence of an instability in a lattice theory usually
becomes obvious when one tries to simulate the theory numerically.
\end{exercise}

We conclude that there exists a set of values for the couplings~$m(a)$,
$\lambda(a)$, $c(a)$, and~$d(a)$ such that the lattice theory gives
the same physical results as the continuum theory up to corrections of
order~$(a\overline{p})^4$ when $\overline{p}\ll\pi/a$. In the classical
limit of our lattice theory, only the $c(a)$ coupling is necessary:
taking $c(a) = 1$ cancels the $\order(a^2)$~errors from the derivative
operator in the original lagrangian. In the quantum theory, $c(a)$ must be
shifted, and a nonzero~$d(a)$ added to cancel
$\order(a^2)$~errors induced by quantum fluctuations. This raises the
problem of how we should determine the correct values for these couplings.
In principle we might run simulations for many different values of the
couplings, and numerically search for the set that gives correct
physics. But a defect of our lattice theory is that it has many more coupling
constants than the continuum theory\,---\,usually far too many to tune
numerically. This is a serious obstacle to the use of improved lagrangians.
In many theories, including QCD, the situation is salvaged by using
perturbation theory to express the extra couplings in terms of the original
two couplings,~$m(a)$ and~$\lambda(a)$. In perturbation theory, we expect
\begin{eqnarray}
c(a) &=& 1 + c_2(a m)\,\lambda(a)^2 + \cdots, \nl
d(a) &=& d_3(a m)\,\lambda(a)^3 + \cdots.
\end{eqnarray}
The radiative corrections for such couplings involve only momenta of
order~$\pi/a$ and larger since these are the momenta where the lattice theory
needs correcting. In asymptotically free theories, like QCD, the fundamental
coupling constant is small and perturbation theory is valid when the lattice
spacing is small. In such situations perturbation theory can be used to
correct the lagrangian, even if the long-distance behavior of the theory is
highly nonperturbative.

To summarize, finite-$a$ errors in quantum lattice
theories are very similar to those for classical lattice theories. In
both cases the errors can be removed, order by order in~$a$, by correcting
the lattice operators. The main difference in the quantum case is that the
coefficients of the correction terms in expressions like \eq{clcorr} are
renormalized by $a$-dependent quantum loop effects that are very specific to
the particular theory and context in which the operator is used.

\begin{exercise}
When the lattice couplings are perturbative, they may be computed by
matching perturbative results from the lattice theory against the
corresponding results generated with the continuum theory. If one has
$n$~couplings to compute, one chooses $n$~low-momentum physical quantities
to match. Each quantity is computed, order by order in perturbation theory,
both in the lattice theory and in the continuum theory. The lattice
couplings are then adjusted so that the lattice results agree with the
continuum results. There are no ultraviolet infinities in the lattice
calculations, since the theory has a finite cutoff, and therefore the bare
couplings are all finite numbers.

The coupling~$d(a)$ is easily computed by examining
the one-loop scattering amplitudes for $\phi\phi\to\phi\phi\phi\phi$; and
the coupling~$c(a)$ can be obtained from the dispersion relation, relating
the energy and momentum of a $\phi$~particle, that is obtained by locating
the poles in the $\phi\to\phi$~propagator. Sketch these calculations to
one-loop order.
\end{exercise}

\subsection{QCD on a Lattice}
Our discussion in the previous section indicates that there are really two
restrictions on the maximum lattice spacing~$a$ that is useful
for numerical work in QCD:
\begin{enumerate}
\item[1)] $a$ must be smaller than any important scale in the hadronic system
under study. Since light hadrons have diameters between 1.5~and 2~fm, the
lattice spacing should probably be no larger than about 0.5~fm. Such a
lattice spacing should suffice for calculations of static properties, like
masses, magnetic moments, charge radii, etc. Smaller lattice spacings are
needed if the hadrons have nonzero momenta.
\item[2)] $a$~must be sufficiently small that perturbation theory works well
at distances smaller than~$a$ (or momenta larger than~$\pi/a$). If this is
the case we can use perturbation theory to compute the correction terms
needed in the lagrangian to obtain high precision from coarse lattices; in
effect, we can fill in the cracks in the lattice.
\end{enumerate}
The second of these restrictions is the more controversial.
There is much
conventional wisdom suggesting that lattice spacings as small as
0.1--0.05~fm might be required to satisfy this condition. I will argue, in
subsequent sections, that $a$'s as large as 0.5~fm are probably alright.
This means that lattices can be much coarser than is usual today.
Before pursuing this crucial point, we must first discuss the
formulation of QCD on a lattice.

The continuum lagrangian for QCD is
\be
\lag = -\half \Tr \F\mu\nu^2
\ee
where
\be
\F\mu\nu \equiv \partial_\mu A_\nu - \partial_\nu A_\mu + ig [A_\mu,A_\nu]
\ee
is the field tensor, a traceless $3\times3$ hermitian matrix. The defining
characteristic of the theory is its invariance with respect to gauge
transformations where
\be
\F\mu\nu \to \Omega(x)\,\F\mu\nu\,\Omega(x)^{-1}
\ee
and $\Omega(x)$ is an arbitrary $x$-dependent \su\ matrix.

The standard discretization of this theory seems perverse at first sight.
Rather than specifying the gauge field by the values of~$A_\mu(x)$ at the
nodes of the lattice, the field is specified by variables on the ``links''
joining the nodes. In the classical theory, the ``link variable'' on
the link joining a node at~$x$ to one at $x+a\hat\mu$ is determined by the
line integral of $A_\mu$ along the link:
\be
\U\mu(x) \equiv {\rm P}\,\exp\left(
-ig\int_x^{x+a\hat\mu} A\cdot dy \right)
\ee
where the P-operator path-orders the $A_\mu$'s along the integration path.
We use~$U_\mu$'s in place of~$A_\mu$'s on
the lattice, because it is {\em impossible\/} to formulate a lattice version
of QCD directly in terms of~$A_\mu$'s that has exact gauge invariance. The
$U_\mu$'s, on the other hand, transform very simply under a gauge
transformation:
\be
\U\mu(x)\to\Omega(x+a\hat\mu)\,\U\mu(x)\,\Omega(x)^{-1}.
\ee
This makes it easy to build a discrete theory with exact gauge invariance.

\begin{exercise}
Show that a gauge invariant object can be associated with any closed path
on the lattice by forming the product of $\U\mu$'s and $\Udag\mu$'s
associated with the links of the path, and taking a trace. In general
$\U\mu(x)$ is associated with a link leaving site~$x$ in (positive)
direction~$\mu$, while $\Udag\mu(x)$ is associated with a link
entering site~$x$ from direction~$\mu$. The trace of a product of link
variables along a closed path is known as a ``Wilson loop.''
\end{exercise}

You might wonder why we go to so much trouble to preserve gauge invariance
when we quite willing give up Lorentz invariance, rotation
invariance, etc. The reason is quite practical. With gauge invariance, the
quark-gluon, three-gluon, and four-gluon couplings in QCD are all equal,
and the bare gluon mass is zero. Without gauge invariance, each of these
couplings must be tuned independently and a gluon mass introduced if one is
to recover QCD. Tuning this many parameters in a numerical simulation is
very expensive. This is not much of a problem in the classical theory,
where approximate gauge invariance keeps the couplings approximately equal;
but it is serious in the quantum theory because quantum fluctuations
(loop-effects) renormalize the various couplings differently in the absence
of exact gauge invariance. So while it is quite possible to formulate
lattice QCD directly in terms of~$\A\mu$'s, the resulting theory would have
only approximate gauge invariance, and thus would be prohibitively expensive
to simulate. Symmetries like Lorentz invariance can be given up with little
cost because the symmetries of the lattice, though far less restrictive, are
still sufficient to prevent the introduction of new interactions with new
couplings (at least to lowest order in~$a$).

We must now build a lattice lagrangian from the link operators. We require
that the lagrangian be gauge invariant, local, and symmetric with respect
to axis interchanges (which is all that is left of Lorentz invariance). The
most local nontrivial gauge invariant object one can build from the link
operators is the ``plaquette operator,'' which involves the product of link
variables around the smallest square at site~$x$ in the $\mu\nu$~plane:
\be
\Tr \Uplaq\mu\nu(x) \equiv \Tr\!\left(\U\mu(x)\U\nu(x+a\hat\mu)
\Udag\mu(x+a\hat\mu+a\hat\nu)\Udag\nu(x)\right).
\ee
To see what this object is, consider evaluating the plaquette centered
about a point~$x_0$ for a very smooth weak classical $\A\mu$~field.
In this limit,
\be \label{uplaqa}
\Tr \Uplaq\mu\nu \approx 3
\ee
since
\be
\U\mu \approx \e^{-iga\A\mu} \approx 1 .
\ee
Given that $\A\mu$~is slowly varying, its value anywhere on the plaquette
should be accurately specified by its value and derivatives at~$x_0$. Thus
the corrections to \eq{uplaqa} should be a polynomial in~$a$ with
coefficients formed from gauge-invariant combinations of~$\A\mu(x_0)$ and
its derivatives. Thus we expect
\begin{eqnarray}
\Tr\Uplaq\mu\nu \,= \, 3
&-&c_1\,a^4\,\Tr\!\left(g\F\mu\nu(x_0)\right)^2 \nl
&-&c_2\,a^6\,\Tr\!\left(g\F\mu\nu(x_0)(D_\mu^2+D_\nu^2)g\F\mu\nu(x_0)
 \right)\nl
&+&\order(a^8)
\label{ope}
\end{eqnarray}
where $c_1$ and $c_2$ are constants, and $D_\mu$ is a
gauge-covariant derivative. The leading correction is order~$a^4$ because
$\F\mu\nu^2$ is the lowest-dimension gauge-invariant combination of
derivatives of $\A\mu$, and it has dimension~4. It is a simple exercise to
show that $c_1 = 1/2$ and $c_2=1/24$ (in the classical limit).

The expansion in \eq{ope} is the classical analogue of an operator product
expansion. It provides the classical relation between the plaquette
operator and the local gauge-invariant operators of the continuum theory.
This relationship is preserved in the quantum theory except that: a)~the
expansion parameters~$c_i$ are renormalized by quantum fluctuations; and
b)~additional terms enter at order~$a^6$ and higher. So the lattice
equivalent of the continuum lagrangian in the quantum theory is
\be
\frac{3}{g^2a^4 u_0^4}\,{\rm Re}\left(\third\Tr\Uplaq\mu\nu - 1\right)
\equiv -\half\Tr\F\mu\nu^2,
\ee
up to corrections of order~$a^2$.
Here $u_0$ accounts for the quantum renormalization of the classical
relation; it is usually absorbed into the definition of the coupling:
\be \label{alpharen}
g_\lat \equiv g\,u_0^2.
\ee
Thus the lattice action for QCD is
\be \label{gluonaction}
S = \beta \sum_{x,\mu\nu} {\rm Re}\left(\third\Tr\Uplaq\mu\nu\right)
   + \mbox{constant}
\ee
where
\be
\beta \equiv  \frac{6}{g_\lat^2} .
\ee
Notice that the lattice spacing has disappeared from the action. It is
customary in lattice simulations to use ``lattice units'' for which~$a=1$,
thereby completely removing~$a$ from the simulation. The lattice spacing
enters only implicitly, through the numerical value of the
parameter~$\beta$; different $\beta$'s correspond to different~$a$'s since
the bare coupling constant is a function of the lattice spacing.

\begin{exercise}
Find correction terms for the classical lattice lagrangian that remove the
order~$a^2$ errors. Consider, for example, adding a term that involves
$a\times 2a$ rectangular Wilson loops.
\end{exercise}

\begin{exercise}
In the classical lattice theory there is only one order~$a^2$ correction.
When quantum effects are included, other operators appear in this order.
What operators are possible? Remember that these operators must be local,
gauge invariant, and respect lattice symmetries.
\end{exercise}

\section{Perturbation Theory: Where are the Limits in Lattice QCD?}
As we discussed in the previous section, the cost of numerical simulations
of QCD is dramatically reduced as the lattice spacing~$a$ is increased. The
resulting loss in accuracy can be minimized by improving the
lattice lagrangian, using perturbation theory to compute the corrections.
Such perturbative improvement is only possible if perturbation theory is
applicable at momenta of order~$\pi/a$. Thus the critical question as we
increase~$a$ is whether or not perturbation theory works at momenta
of~$\pi/a$ and larger.

Empirical evidence suggests that perturbative QCD applied to continuum
quantities works down to momenta as low as~1~GeV; for example, $\tau$ decay
provides one of the most accurate determinations of the strong coupling
constant. Thus lattice spacings as large as~0.5~fm ought to
work. This result is very encouraging, but it is important to confirm it
with detailed numerical tests using lattice simulations. Indeed, given the
tremendous control we have over simulations, it is far easier to test
perturbative QCD against nonperturbative simulations than against
experimental data.

One common procedure for testing perturbation theory using simulations is
to compute the $\beta$~function\,---\,a perturbative quantity\,---\,by
comparing values for some physical quantity, like the  glueball mass or the
string tension, computed at various values of the bare coupling. This
gives a nonperturbative determination of the $\beta$~function to
compare with the perturbative prediction. In my opinion, this is a {\em
bad\/} way to test perturbation theory since failure in such a test could
reflect one of two possible problems: either perturbation theory is not
working, or there are large $\order(a,a^2)$~errors in the calculation of
the physical quantity used to determine the $\beta$~function. It is
fairly difficult to rule out the second option numerically, and so a
negative result is ambiguous.

A better procedure is to follow phenomenologists, who identify
short-distance quantities that can be both measured nonperturbatively and
calculated in perturbation theory. As lattice theorists, we have a much
simpler job than the phenomenologist since we can easily design dozens of
quantities that are very ultraviolet and that are easy to measure in
simulations; we don't have to worry about experimental cuts, Monte Carlo
corrections for hadronization, and all the other problems associated with
experimental data. The idea is to compute various short-distance
quantities\,---\,for example, vacuum matrix elements of local operators
like~$\Tr\Uplaq\mu\nu$\,---\,using numerical simulations and then to
compare the results with predictions from perturbation theory.

Careful testing of perturbation theory requires a careful definition of the
expansion parameter, the strong coupling constant. This is as true in
lattice calculations as it is in continuum calculations. There are two
aspects to the problem. First we need to define what we mean by the running
coupling constant~$\alpha_s(q)$ (``the scheme''), and then we need a
procedure for choosing a~$q$ (``the scale'') appropriate to the quantity of
interest.  We deal with these issues in the next section. Then, in the
following section, we compare perturbative results with results from
nonperturbative simulations. Note that our choice of scheme and scale
for~$\alpha_s$ is completely automatic; there is no fine tuning of the
perturbative results we use in our comparison with simulations.

\subsection{Defining $\alpha_s$} \label{definingalpha}
Given one definition~$\alpha_1(q)$ it is easy to create
other definitions: for example, $\alpha_2(q)\equiv\alpha_1(2q)$ or
$\alpha_3(q)\equiv\alpha_1(q)+3\alpha_1(q)^2+\cdots$. In principle, any of
these can be used as the expansion parameter for perturbation theory; in
practice some definitions will be more reliable than others. For example,
if $\alpha_1(q)$ is a reliable choice then
$\alpha_4(q)\equiv\alpha_1(q)+100\alpha_1(q)^2$ will probably be useless.
Expansions in terms of either coupling will be the same in lowest order,
but higher orders will have large negative corrections when $\alpha_4$ is
used. To minimize such problems, I like to use a physical quantity to
define the coupling. One such definition that is particularly convenient
uses the potential between a static quark and antiquark:
\be
V(q) \equiv -\,\frac{C_F\,4\pi\,\alpha_V(q)}{q^2}
\ee
where $C_F = 4/3$ and $q$ is the momentum transferred between the quark and
antiquark. I like this definition because there is no ambiguity about the
relationship between the argument of~$\alpha_V$ and the momentum scale in
the process:
\be
\alpha_V(q) = \parbox{2in}{the coupling strength of a \\ gluon with
momentum~$q$.}
\ee
This is a useful definition both for continuum and for lattice results. It
is easy to convert expansions in terms of $\alpha_\msb$ to $\alpha_V$
expansions by using\footnote{In practice there is little difference between
using~$\alpha_\msb$ or
using~$\alpha_V$, but the latter suggests a simple procedure for setting the
scale.}
 \be
\alpha_\msb(q) = \alpha_V(\e^{5/6} q) \left\{ 1 + 2\alpha_V/\pi
 +\cdots \right\}.
 \ee

Having settled on~$\alpha_V(q)$, we now need some way of specifying the
scale~$q$.
The procedure is best understood through an example. Consider the vacuum
expectation value~$\avg{1-\third\Tr\U\mu}$ computed in Landau gauge. To
leading order, this quantity is proportional to $\avg{\Tr\A\mu^2}$ and
the leading contribution in lattice perturbation theory is
\be
\avg{1-\third\Tr\U\mu}_{LG} =
\alpha_V(q^*)\,{2\pi}\,{a^2}\,\int_{\pi/a}^{-\pi/a}
\frac{d^4q}{(2\pi)^4}\,\frac{1}{\hat{q}^2} +\order(\alpha_V^2)
\ee
where $\hat{q}_\mu \equiv (2/a)\sin(a\,q_\mu/2)$, and $1/\hat{q}^2$ comes
from a gluon propagator. The question is what values for~$q^*$ makes sense.
Given that~$\alpha_V$ is the coupling for a gluon of momentum~$q$, we
really want
\be \label{qstint}
\alpha_V(q^*) \int_{\pi/a}^{-\pi/a}
\frac{d^4q}{\hat{q}^2}
 \equiv
 \int_{\pi/a}^{-\pi/a}
\frac{d^4q}{\hat{q}^2}\,\alpha_V(q)
\ee
except that the right-hand side diverges at small $q$. To see why it
diverges, we can rewrite that side of the equation in terms of
$\alpha_V(\mu)$ for some fixed value of~$\mu$:
\be
 \int_{-\pi/a}^{\pi/a}
\frac{d^4q}{\hat{q}^2}\,\left\{ \alpha_V(\mu)
 - \beta_0\alpha_V^2(\mu)\ln(q^2/\mu^2) -
 \beta_0^2\alpha_V^3(\mu)\ln^2(q^2/\mu^2) - \cdots\right\}.
 \ee
Each term in this expansion separately is finite, but the sum diverges.
This is not surprising since perturbation theory yields only an asymptotic
expansion. Furthermore it is inconsistent to keep an infinite number of
radiative corrections in this expansion while only keeping only one or two
orders elsewhere in a calculation. So, for present purposes, we should
replace the $\alpha_V$'s in \eq{qstint} by the first two terms
in their expansion in terms of~$\alpha_V(\mu)$. It is then easy to solve
for $q^*$ to obtain:
\be
\ln(q^*) = \,\frac{\displaystyle
\int_{-\pi/a}^{\pi/a}\frac{d^4q}{\hat{q}^2}\,\ln(q)}{\displaystyle
 \int_{-\pi/a}^{\pi/a}\frac{d^4q}{\hat{q}^2}}\,.
\ee
Evaluating the integrals one obtains $q^*=2.8/a$, and a final prediction of
\be
\avg{1-\third\Tr\U\mu}_{LG} = 0.97\,\alpha_V(2.8/a)
\ee
The analogous quantity in the continuum has a quadratic ultraviolet
divergence, and so it is not surprising that the average momentum
scale~$q^*$ in this process very nearly equals the maximum momentum
allowed on the lattice.

This example illustrates a general procedure. A quantity whose first order
contribution has the form
\be
\alpha_V(q^*)\,\int_{-\pi/a}^{\pi/a}d^4q\,f(q),
\ee
where $q$ is the gluon momentum, has scale\footnote{This definition needs
modification if there are nearly canceling contributions coming from
very different scales. Luckily this
situation does not arise often, and not at all in the applications we are
considering here.}
\be
\ln(q^*) =\, \frac{\displaystyle\int_{-\pi/a}^{\pi/a} d^4q\,f(q)\,\ln(q)}{
\displaystyle\int_{-\pi/a}^{\pi/a}d^4q\,f(q)} \,.
\ee
Each perturbative quantity has its own characteristic scale; the more
ultraviolet a quantity is, the larger is its~$q^*$. And, in an
asymptotically free theory, the larger~$q^*$ is, the more accurate
is perturbation theory.

\subsection{Testing Perturbation Theory}
To probe the validity of perturbation theory, we need simulation results
for several short-distance quantities and we need perturbative
predictions for those quantities. Here we focus on logarithms of the
expectation values of planar
$m\times n$~Wilson loops,
\be
\ln W_{mn} \equiv \ln\avg{\third\Tr\U{m\times n}},
\ee
because these are easy to simulate and have been analyzed in perturbation
theory through second order. (We use the logarithms of the loops
rather than the loops themselves because the logarithms have more convergent
expansions\,---\,ultraviolet self-energies proportional to the length of the
loops exponentiate and so are better handled by taking a logarithm.)
Small Wilson loops are among the most ultraviolet quantities there are in
lattice gauge theory. To probe lower momenta, we examine  ``Creutz
ratios'' of the loop values,
\be
\chi_{mn} \equiv -\ln\left(\frac{W_{mn} W_{m-1\,n-1}}{W_{m\,n-1}W_{m-1\,n}}
\right).
\ee
The most ultraviolet parts of the loops cancel in Creutz ratios and so these
tend to be more infrared than the $W$'s (ie, have lower~$q^*$'s). In
general quantities with larger  loops and/or smaller~$q^*$'s are more
likely to have significant nonperturbative contributions.

To make perturbative predictions we must determine~$\alpha_V$ for
each value of~$\beta$ used in the gluon action~\eq{gluonaction} (ie, for each
value of the lattice spacing). In principle we could do this
by measuring the static-quark potential in the simulation. However it is
simpler to redefine the coupling in terms of the expectation value of the
plaquette operator. Using perturbation theory, one finds that
\be \label{ptplaq}
-\ln W_{11} = \frac{4\pi}{3}\,\alpha_V(3.4/a)\,\left\{1 - (1.19+0.02
n_f)\,\alpha_V + \order(\alpha_V^2)\right\},
\ee
where scale~$q^*=3.4/a$ is determined using the methods of the
previous section, and $n_f$~is the number of light quark flavors
in the simulation ($n_f=0$ for the results we show here). This relation can
be taken as a definition of $\alpha_V$. The coupling is determined
nonperturbatively by inverting the series to determine $\alpha_V(3.4/a)$
from simulated values for $W_{11}$. Having a value for~$\alpha_V(3.4/a)$,
we obtain~$\alpha_V(q)$ at any other scale using two-loop  perturbative
evolution:
\begin{eqnarray}
\alpha_V^{-1}(q) &=& \beta_0\ln(q^2/\Lambda_V^2) +
\beta_1/\beta_0\ln\ln(q^2/\Lambda_V^2) + \order(\alpha_V) \nl
&\approx& \alpha_V^{-1}(q_0) +
\left(\beta_0+\beta_1\alpha_V(q_0)\right)\,\ln(q^2/q^2_0)
\end{eqnarray}
where  $\beta_0 = (11-\frac{2}{3}n_f)/4\pi$,
$\beta_1 = (102-\frac{38}{3}n_f)/16\pi^2$, and $\Lambda_V$ is a constant
(the ``scale parameter'').

Simulation results for~$\alpha_V$ at several
different~$\beta$'s are given in Table~\ref{alphas}.
Values for $a\Lambda_V$ are also listed. Since~$\Lambda_V$ is independent
of~$\beta$, these last numbers can be used to compute ratios of the lattice
spacings at different~$\beta$'s. The inverse lattice spacing at~$\beta=5.7$
is about 1~GeV so that $a\approx0.2$~fm.
\begin{table} \centering
\begin{tabular}{c|ccccc}
$\beta$ & $-\ln W_{11}$ & $\alpha_V(1/a)$ & $\alpha_V(\pi/a)$ &
$a\,\Lambda_V$ \\ \hline
5.4 & 0.7516  & $\approx 1$ & 0.271 & 0.627 \\
5.7 & 0.5995  & 0.355  & 0.188 & 0.293\\
6.0 & 0.5214  & 0.247    & 0.156 & 0.169\\
6.2 & 0.4884  & 0.214  & 0.143 & 0.127\\
6.4 & 0.4610  & 0.191  & 0.133 & 0.097\\
6.8 & 0.4167  & 0.160  & 0.117 & 0.058\\
9.0 & 0.2795   & 0.088    & 0.074 & $4.1\times10^{-3}$\\
12 & 0.1954  & 0.056   & 0.050 & $1.2\times10^{-4}$ \\
\end{tabular}
\caption{ Monte Carlo data for logarithm of the plaquette, together with the
coupling constant values and the~$\alpha_V$ scale parameter in lattice
units. The inverse lattice spacing at~$\beta=5.7$ is $a^{-1}\approx 1$~GeV.
All results are for the standard lattice action with $n_f=0$.}
\label{alphas}
\end{table}

\begin{figure} \centering
\setlength{\unitlength}{0.240900pt}
\ifx\plotpoint\undefined\newsavebox{\plotpoint}\fi
\sbox{\plotpoint}{\rule[-0.175pt]{0.350pt}{0.350pt}}%
\begin{picture}(1500,900)(0,0)
\tenrm
\sbox{\plotpoint}{\rule[-0.175pt]{0.350pt}{0.350pt}}%
\put(264,158){\rule[-0.175pt]{282.335pt}{0.350pt}}
\put(264,158){\rule[-0.175pt]{4.818pt}{0.350pt}}
\put(242,158){\makebox(0,0)[r]{$0$}}
\put(1416,158){\rule[-0.175pt]{4.818pt}{0.350pt}}
\put(264,284){\rule[-0.175pt]{4.818pt}{0.350pt}}
\put(242,284){\makebox(0,0)[r]{$0.1$}}
\put(1416,284){\rule[-0.175pt]{4.818pt}{0.350pt}}
\put(264,410){\rule[-0.175pt]{4.818pt}{0.350pt}}
\put(242,410){\makebox(0,0)[r]{$0.2$}}
\put(1416,410){\rule[-0.175pt]{4.818pt}{0.350pt}}
\put(264,535){\rule[-0.175pt]{4.818pt}{0.350pt}}
\put(242,535){\makebox(0,0)[r]{$0.3$}}
\put(1416,535){\rule[-0.175pt]{4.818pt}{0.350pt}}
\put(264,661){\rule[-0.175pt]{4.818pt}{0.350pt}}
\put(242,661){\makebox(0,0)[r]{$0.4$}}
\put(1416,661){\rule[-0.175pt]{4.818pt}{0.350pt}}
\put(411,158){\rule[-0.175pt]{0.350pt}{4.818pt}}
\put(411,113){\makebox(0,0){$6$}}
\put(411,767){\rule[-0.175pt]{0.350pt}{4.818pt}}
\put(557,158){\rule[-0.175pt]{0.350pt}{4.818pt}}
\put(557,113){\makebox(0,0){$7$}}
\put(557,767){\rule[-0.175pt]{0.350pt}{4.818pt}}
\put(704,158){\rule[-0.175pt]{0.350pt}{4.818pt}}
\put(704,113){\makebox(0,0){$8$}}
\put(704,767){\rule[-0.175pt]{0.350pt}{4.818pt}}
\put(850,158){\rule[-0.175pt]{0.350pt}{4.818pt}}
\put(850,113){\makebox(0,0){$9$}}
\put(850,767){\rule[-0.175pt]{0.350pt}{4.818pt}}
\put(997,158){\rule[-0.175pt]{0.350pt}{4.818pt}}
\put(997,113){\makebox(0,0){$10$}}
\put(997,767){\rule[-0.175pt]{0.350pt}{4.818pt}}
\put(1143,158){\rule[-0.175pt]{0.350pt}{4.818pt}}
\put(1143,113){\makebox(0,0){$11$}}
\put(1143,767){\rule[-0.175pt]{0.350pt}{4.818pt}}
\put(1290,158){\rule[-0.175pt]{0.350pt}{4.818pt}}
\put(1290,113){\makebox(0,0){$12$}}
\put(1290,767){\rule[-0.175pt]{0.350pt}{4.818pt}}
\put(264,158){\rule[-0.175pt]{282.335pt}{0.350pt}}
\put(1436,158){\rule[-0.175pt]{0.350pt}{151.526pt}}
\put(264,787){\rule[-0.175pt]{282.335pt}{0.350pt}}
\put(67,472){\makebox(0,0)[l]{\shortstack{$\chi_{22}$}}}
\put(850,68){\makebox(0,0){$\beta = 6/g^2_{\rm lat}$}}
\put(337,724){\makebox(0,0)[l]{$1^{\rm st}$ Order Perturbation Theory}}
\put(264,158){\rule[-0.175pt]{0.350pt}{151.526pt}}
\put(1306,722){\makebox(0,0)[r]{Monte Carlo}}
\put(1350,722){\circle{24}}
\put(367,628){\circle{24}}
\put(411,492){\circle{24}}
\put(440,451){\circle{24}}
\put(850,286){\circle{24}}
\put(1290,241){\circle{24}}
\put(1306,677){\makebox(0,0)[r]{$\alpha_{V}(q^*)$}}
\put(1350,677){\raisebox{-1.2pt}{\makebox(0,0){$\Diamond$}}}
\put(367,662){\raisebox{-1.2pt}{\makebox(0,0){$\Diamond$}}}
\put(411,517){\raisebox{-1.2pt}{\makebox(0,0){$\Diamond$}}}
\put(440,472){\raisebox{-1.2pt}{\makebox(0,0){$\Diamond$}}}
\put(850,290){\raisebox{-1.2pt}{\makebox(0,0){$\Diamond$}}}
\put(1290,242){\raisebox{-1.2pt}{\makebox(0,0){$\Diamond$}}}
\put(1306,632){\makebox(0,0)[r]{$\alpha_{\rm \overline{MS}}(q^*)$}}
\put(1350,632){\raisebox{-1.2pt}{\makebox(0,0){$\Box$}}}
\put(367,533){\raisebox{-1.2pt}{\makebox(0,0){$\Box$}}}
\put(411,449){\raisebox{-1.2pt}{\makebox(0,0){$\Box$}}}
\put(440,420){\raisebox{-1.2pt}{\makebox(0,0){$\Box$}}}
\put(850,281){\raisebox{-1.2pt}{\makebox(0,0){$\Box$}}}
\put(1290,238){\raisebox{-1.2pt}{\makebox(0,0){$\Box$}}}
\put(1306,587){\makebox(0,0)[r]{$\alpha_{\rm lat}$}}
\put(1350,587){\makebox(0,0){$\times$}}
\put(367,286){\makebox(0,0){$\times$}}
\put(411,279){\makebox(0,0){$\times$}}
\put(440,276){\makebox(0,0){$\times$}}
\put(850,239){\makebox(0,0){$\times$}}
\put(1290,219){\makebox(0,0){$\times$}}
\end{picture}
 \\
\setlength{\unitlength}{0.240900pt}
\ifx\plotpoint\undefined\newsavebox{\plotpoint}\fi
\sbox{\plotpoint}{\rule[-0.175pt]{0.350pt}{0.350pt}}%
\begin{picture}(1500,900)(0,0)
\tenrm
\sbox{\plotpoint}{\rule[-0.175pt]{0.350pt}{0.350pt}}%
\put(264,158){\rule[-0.175pt]{282.335pt}{0.350pt}}
\put(264,158){\rule[-0.175pt]{4.818pt}{0.350pt}}
\put(242,158){\makebox(0,0)[r]{$0$}}
\put(1416,158){\rule[-0.175pt]{4.818pt}{0.350pt}}
\put(264,284){\rule[-0.175pt]{4.818pt}{0.350pt}}
\put(242,284){\makebox(0,0)[r]{$0.1$}}
\put(1416,284){\rule[-0.175pt]{4.818pt}{0.350pt}}
\put(264,410){\rule[-0.175pt]{4.818pt}{0.350pt}}
\put(242,410){\makebox(0,0)[r]{$0.2$}}
\put(1416,410){\rule[-0.175pt]{4.818pt}{0.350pt}}
\put(264,535){\rule[-0.175pt]{4.818pt}{0.350pt}}
\put(242,535){\makebox(0,0)[r]{$0.3$}}
\put(1416,535){\rule[-0.175pt]{4.818pt}{0.350pt}}
\put(264,661){\rule[-0.175pt]{4.818pt}{0.350pt}}
\put(242,661){\makebox(0,0)[r]{$0.4$}}
\put(1416,661){\rule[-0.175pt]{4.818pt}{0.350pt}}
\put(411,158){\rule[-0.175pt]{0.350pt}{4.818pt}}
\put(411,113){\makebox(0,0){$6$}}
\put(411,767){\rule[-0.175pt]{0.350pt}{4.818pt}}
\put(557,158){\rule[-0.175pt]{0.350pt}{4.818pt}}
\put(557,113){\makebox(0,0){$7$}}
\put(557,767){\rule[-0.175pt]{0.350pt}{4.818pt}}
\put(704,158){\rule[-0.175pt]{0.350pt}{4.818pt}}
\put(704,113){\makebox(0,0){$8$}}
\put(704,767){\rule[-0.175pt]{0.350pt}{4.818pt}}
\put(850,158){\rule[-0.175pt]{0.350pt}{4.818pt}}
\put(850,113){\makebox(0,0){$9$}}
\put(850,767){\rule[-0.175pt]{0.350pt}{4.818pt}}
\put(997,158){\rule[-0.175pt]{0.350pt}{4.818pt}}
\put(997,113){\makebox(0,0){$10$}}
\put(997,767){\rule[-0.175pt]{0.350pt}{4.818pt}}
\put(1143,158){\rule[-0.175pt]{0.350pt}{4.818pt}}
\put(1143,113){\makebox(0,0){$11$}}
\put(1143,767){\rule[-0.175pt]{0.350pt}{4.818pt}}
\put(1290,158){\rule[-0.175pt]{0.350pt}{4.818pt}}
\put(1290,113){\makebox(0,0){$12$}}
\put(1290,767){\rule[-0.175pt]{0.350pt}{4.818pt}}
\put(264,158){\rule[-0.175pt]{282.335pt}{0.350pt}}
\put(1436,158){\rule[-0.175pt]{0.350pt}{151.526pt}}
\put(264,787){\rule[-0.175pt]{282.335pt}{0.350pt}}
\put(67,472){\makebox(0,0)[l]{\shortstack{$\chi_{22}$}}}
\put(850,68){\makebox(0,0){$\beta = 6/g^2_{\rm lat}$}}
\put(337,724){\makebox(0,0)[l]{$2^{\rm nd}$ Order Perturbation Theory}}
\put(264,158){\rule[-0.175pt]{0.350pt}{151.526pt}}
\put(1306,722){\makebox(0,0)[r]{Monte Carlo}}
\put(1350,722){\circle{24}}
\put(367,628){\circle{24}}
\put(411,492){\circle{24}}
\put(440,451){\circle{24}}
\put(850,286){\circle{24}}
\put(1290,241){\circle{24}}
\put(1306,677){\makebox(0,0)[r]{$\alpha_{V}(q^*)$}}
\put(1350,677){\raisebox{-1.2pt}{\makebox(0,0){$\Diamond$}}}
\put(367,603){\raisebox{-1.2pt}{\makebox(0,0){$\Diamond$}}}
\put(411,487){\raisebox{-1.2pt}{\makebox(0,0){$\Diamond$}}}
\put(440,449){\raisebox{-1.2pt}{\makebox(0,0){$\Diamond$}}}
\put(850,286){\raisebox{-1.2pt}{\makebox(0,0){$\Diamond$}}}
\put(1290,241){\raisebox{-1.2pt}{\makebox(0,0){$\Diamond$}}}
\put(1306,632){\makebox(0,0)[r]{$\alpha_{\rm \overline{MS}}(q^*)$}}
\put(1350,632){\raisebox{-1.2pt}{\makebox(0,0){$\Box$}}}
\put(367,575){\raisebox{-1.2pt}{\makebox(0,0){$\Box$}}}
\put(411,475){\raisebox{-1.2pt}{\makebox(0,0){$\Box$}}}
\put(440,441){\raisebox{-1.2pt}{\makebox(0,0){$\Box$}}}
\put(850,285){\raisebox{-1.2pt}{\makebox(0,0){$\Box$}}}
\put(1290,240){\raisebox{-1.2pt}{\makebox(0,0){$\Box$}}}
\put(1306,587){\makebox(0,0)[r]{$\alpha_{\rm lat}$}}
\put(1350,587){\makebox(0,0){$\times$}}
\put(367,352){\makebox(0,0){$\times$}}
\put(411,339){\makebox(0,0){$\times$}}
\put(440,332){\makebox(0,0){$\times$}}
\put(850,266){\makebox(0,0){$\times$}}
\put(1290,234){\makebox(0,0){$\times$}}
\end{picture}
\caption{
Results for Creutz ratio~$\chi_{22}$  at
different couplings~$\beta$ from  Monte Carlo simulations (circles), and from
perturbation theory (using $\alpha_{V}(q^*)$ (diamonds),
$\alpha_{\msb}(q^*)$ (boxes),  and $\alpha_{\lat}$ (crosses)). The first plot
shows perturbation theory through one-loop order, and the second through
two-loop
order. Statistical errors in the Monte Carlo results are negligible.
\label{chi22}
}
\end{figure}
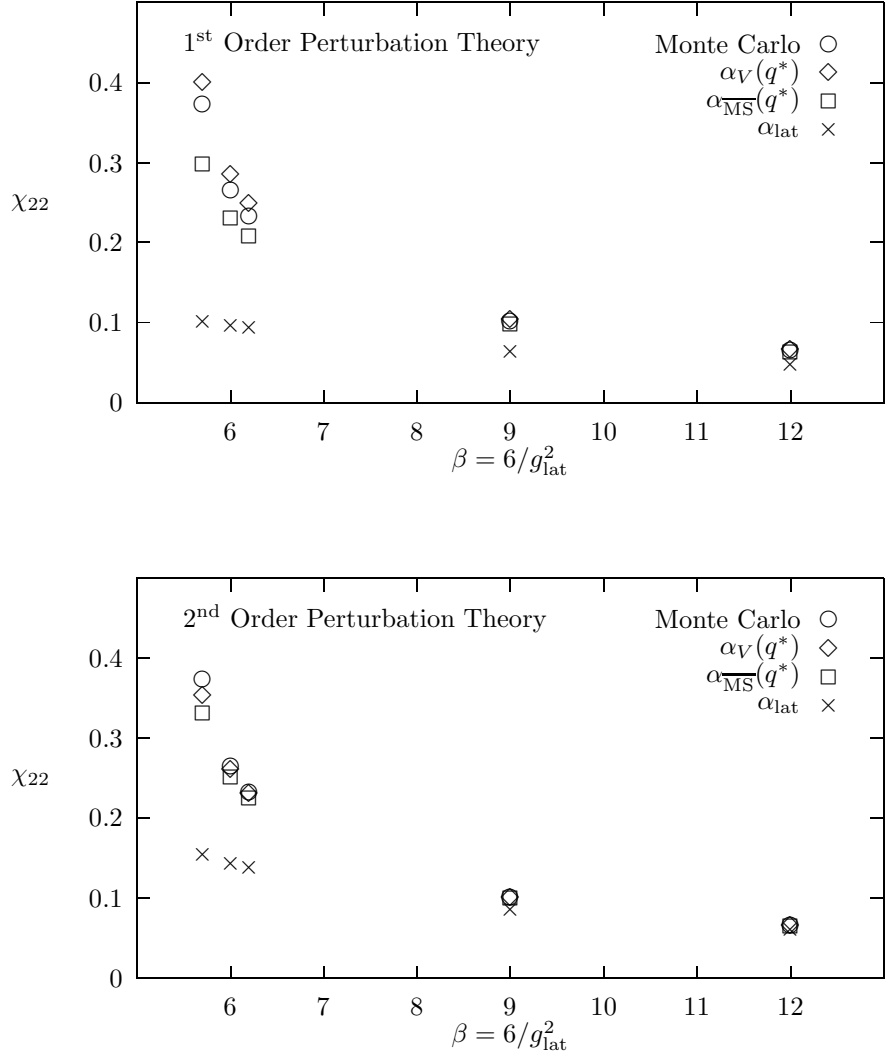

As a first test of perturbation theory, consider the Creutz
ratio~$\chi_{22}$. It has the perturbative expansion
\be
\chi_{22} = 1.2\,\alpha_V(1.1/a) - 0.4\,\alpha_V^2 + \order(\alpha_V^3)
\ee
where, as expected, its scale~$q^* = 1.1/a$ is significantly smaller than
that for the plaquette. Monte Carlo simulation results are compared with
first-order and second-order perturbative predictions for several
$\beta$'s in Figure~\ref{chi22}. Perturbation theory gives excellent results
throughout the range shown.

For comparison results using~$\alpha_\msb(q^*)$
in place of~$\alpha_V(q^*)$ are also shown; they are roughly comparable.
Results are also shown for perturbation theory in terms of the bare
coupling, $\alpha_\lat \equiv g^2_\lat/4\pi$. These are terrible. I show
these last results only because it is a common practice in lattice
perturbation theory to expand in terms of the bare coupling. This is a
serious mistake\,---\,one that has in large measure crippled earlier
attempts at using lattice perturbation theory. Using
$\alpha_\lat$ is wrong for two reasons. First, it ignores the fact that the
Creutz ratio is significantly less ultraviolet than most Wilson loops and so
should be expressed in terms of the running coupling evaluated at a
smaller scale. Second, it ignores the renormalization in the relation
between the lattice coupling and the continuum coupling (\eq{alpharen}),
and, as we discuss in the next section, this renormalization is large.
Expansions in~$\alpha_\lat$ almost always underestimate perturbative
effects.

Our choice of the plaquette for determining~$\alpha_V$ is for convenience.
In principle one ought to be able to use any short-distance quantity to
measure~$\alpha_V$ in a simulation. If perturbation theory is working, the
results obtained from different quantities at a given~$\beta$ should agree.
In Table~\ref{diffalpha} we list numerical values of~$\alpha_V(\pi/a)$ as
determined from a variety of small Wilson loops at several values
of~$\beta$. In each case~$\alpha_V(q^*)$ is adjusted so that the
three-loop peturbative prediction agrees with simulation results, and then
$\alpha_V$ is evolved to scale~$q=\pi/a$ using the two-loop beta function.
The agreement between different determinations of~$\alpha_V$ is good (even
excellent for the smallest loops) for all~$\beta$'s listed,
including~$\beta=5.4$ where the lattice spacing is almost~1/2~fm. (Larger
loops give poorer results because they are more strongly affected by
nonperturbative effects.)
\begin{table}
\begin{center}
\begin{tabular}{c|cccc}
$\alpha_V(\pi/a)$ &$\beta=5.4$ & $\beta=5.7$ & $\beta=6.0$ & $\beta=6.2$ \\
\hline
$-\ln W_{11}$ & 0.271 & 0.188 & 0.156 & 0.143 \\
$-\ln W_{12}$ & 0.269 & 0.190 & 0.156 & 0.143 \\
$-\ln W_{22}$ & 0.271 & 0.194 & 0.156 & 0.143 \\
$-\ln W_{33}$ & 0.293 & 0.205 & 0.159 & 0.145 \\
\end{tabular}
\end{center}
\caption{Values of~$\alpha_V(\pi/a)$ as determined by comparing three-loop
perturbation theory with nonperturbative simulation results for the
expectation values of small planar Wilson loops.}
\label{diffalpha}
\end{table}

\begin{figure}\centering
\hspace{-1in}\parbox{1.7in}{
\setlength{\unitlength}{0.240900pt}
\ifx\plotpoint\undefined\newsavebox{\plotpoint}\fi
\sbox{\plotpoint}{\rule[-0.175pt]{0.350pt}{0.350pt}}%
\begin{picture}(674,1259)(0,0)
\tenrm
\sbox{\plotpoint}{\rule[-0.175pt]{0.350pt}{0.350pt}}%
\put(264,158){\rule[-0.175pt]{83.351pt}{0.350pt}}
\put(264,158){\rule[-0.175pt]{0.350pt}{238.009pt}}
\put(264,440){\rule[-0.175pt]{4.818pt}{0.350pt}}
\put(242,440){\makebox(0,0)[r]{$0.2$}}
\put(590,440){\rule[-0.175pt]{4.818pt}{0.350pt}}
\put(264,723){\rule[-0.175pt]{4.818pt}{0.350pt}}
\put(242,723){\makebox(0,0)[r]{$0.4$}}
\put(590,723){\rule[-0.175pt]{4.818pt}{0.350pt}}
\put(264,1005){\rule[-0.175pt]{4.818pt}{0.350pt}}
\put(242,1005){\makebox(0,0)[r]{$0.6$}}
\put(590,1005){\rule[-0.175pt]{4.818pt}{0.350pt}}
\put(351,158){\rule[-0.175pt]{0.350pt}{4.818pt}}
\put(351,113){\makebox(0,0){$1$}}
\put(351,1126){\rule[-0.175pt]{0.350pt}{4.818pt}}
\put(437,158){\rule[-0.175pt]{0.350pt}{4.818pt}}
\put(437,113){\makebox(0,0){$2$}}
\put(437,1126){\rule[-0.175pt]{0.350pt}{4.818pt}}
\put(524,158){\rule[-0.175pt]{0.350pt}{4.818pt}}
\put(524,113){\makebox(0,0){$3$}}
\put(524,1126){\rule[-0.175pt]{0.350pt}{4.818pt}}
\put(264,158){\rule[-0.175pt]{83.351pt}{0.350pt}}
\put(610,158){\rule[-0.175pt]{0.350pt}{238.009pt}}
\put(264,1146){\rule[-0.175pt]{83.351pt}{0.350pt}}
\put(437,68){\makebox(0,0){$q\,a$}}
\put(134,1146){\makebox(0,0)[l]{$\alpha_V(q)$}}
\put(437,1064){\makebox(0,0)[l]{$\beta=5.7$}}
\put(264,158){\rule[-0.175pt]{0.350pt}{238.009pt}}
\put(559,416){\circle{24}}
\put(530,435){\circle{24}}
\put(524,440){\circle{24}}
\put(493,471){\circle{24}}
\put(485,488){\circle{24}}
\put(477,512){\circle{24}}
\put(358,656){\circle{24}}
\put(359,708){\circle{24}}
\put(348,848){\circle{24}}
\put(357,792){\circle{24}}
\put(333,996){\circle{24}}
\sbox{\plotpoint}{\rule[-0.350pt]{0.700pt}{0.700pt}}%
\put(329,1082){\usebox{\plotpoint}}
\put(329,1054){\rule[-0.350pt]{0.700pt}{6.625pt}}
\put(330,1027){\rule[-0.350pt]{0.700pt}{6.625pt}}
\put(331,999){\rule[-0.350pt]{0.700pt}{6.625pt}}
\put(332,972){\rule[-0.350pt]{0.700pt}{6.625pt}}
\put(333,958){\rule[-0.350pt]{0.700pt}{3.266pt}}
\put(334,944){\rule[-0.350pt]{0.700pt}{3.266pt}}
\put(335,931){\rule[-0.350pt]{0.700pt}{3.266pt}}
\put(336,917){\rule[-0.350pt]{0.700pt}{3.266pt}}
\put(337,904){\rule[-0.350pt]{0.700pt}{3.266pt}}
\put(338,890){\rule[-0.350pt]{0.700pt}{3.266pt}}
\put(339,877){\rule[-0.350pt]{0.700pt}{3.266pt}}
\put(340,863){\rule[-0.350pt]{0.700pt}{3.266pt}}
\put(341,850){\rule[-0.350pt]{0.700pt}{3.266pt}}
\put(342,841){\rule[-0.350pt]{0.700pt}{2.115pt}}
\put(343,832){\rule[-0.350pt]{0.700pt}{2.115pt}}
\put(344,823){\rule[-0.350pt]{0.700pt}{2.115pt}}
\put(345,814){\rule[-0.350pt]{0.700pt}{2.115pt}}
\put(346,806){\rule[-0.350pt]{0.700pt}{2.115pt}}
\put(347,797){\rule[-0.350pt]{0.700pt}{2.115pt}}
\put(348,788){\rule[-0.350pt]{0.700pt}{2.115pt}}
\put(349,779){\rule[-0.350pt]{0.700pt}{2.115pt}}
\put(350,771){\rule[-0.350pt]{0.700pt}{2.115pt}}
\put(351,765){\rule[-0.350pt]{0.700pt}{1.331pt}}
\put(352,759){\rule[-0.350pt]{0.700pt}{1.331pt}}
\put(353,754){\rule[-0.350pt]{0.700pt}{1.331pt}}
\put(354,748){\rule[-0.350pt]{0.700pt}{1.331pt}}
\put(355,743){\rule[-0.350pt]{0.700pt}{1.331pt}}
\put(356,737){\rule[-0.350pt]{0.700pt}{1.331pt}}
\put(357,732){\rule[-0.350pt]{0.700pt}{1.331pt}}
\put(358,726){\rule[-0.350pt]{0.700pt}{1.331pt}}
\put(359,721){\rule[-0.350pt]{0.700pt}{1.331pt}}
\put(360,715){\rule[-0.350pt]{0.700pt}{1.331pt}}
\put(361,710){\rule[-0.350pt]{0.700pt}{1.331pt}}
\put(362,704){\rule[-0.350pt]{0.700pt}{1.331pt}}
\put(363,699){\rule[-0.350pt]{0.700pt}{1.331pt}}
\put(364,693){\rule[-0.350pt]{0.700pt}{1.331pt}}
\put(365,688){\rule[-0.350pt]{0.700pt}{1.331pt}}
\put(366,682){\rule[-0.350pt]{0.700pt}{1.331pt}}
\put(367,677){\rule[-0.350pt]{0.700pt}{1.331pt}}
\put(368,671){\rule[-0.350pt]{0.700pt}{1.331pt}}
\put(369,666){\rule[-0.350pt]{0.700pt}{1.331pt}}
\put(370,660){\rule[-0.350pt]{0.700pt}{1.331pt}}
\put(371,655){\rule[-0.350pt]{0.700pt}{1.331pt}}
\put(372,652){\rule[-0.350pt]{0.700pt}{0.712pt}}
\put(373,649){\rule[-0.350pt]{0.700pt}{0.712pt}}
\put(374,646){\rule[-0.350pt]{0.700pt}{0.712pt}}
\put(375,643){\rule[-0.350pt]{0.700pt}{0.712pt}}
\put(376,640){\rule[-0.350pt]{0.700pt}{0.712pt}}
\put(377,637){\rule[-0.350pt]{0.700pt}{0.712pt}}
\put(378,634){\rule[-0.350pt]{0.700pt}{0.712pt}}
\put(379,631){\rule[-0.350pt]{0.700pt}{0.712pt}}
\put(380,628){\rule[-0.350pt]{0.700pt}{0.712pt}}
\put(381,625){\rule[-0.350pt]{0.700pt}{0.712pt}}
\put(382,622){\rule[-0.350pt]{0.700pt}{0.712pt}}
\put(383,619){\rule[-0.350pt]{0.700pt}{0.712pt}}
\put(384,616){\rule[-0.350pt]{0.700pt}{0.712pt}}
\put(385,613){\rule[-0.350pt]{0.700pt}{0.712pt}}
\put(386,610){\rule[-0.350pt]{0.700pt}{0.712pt}}
\put(387,607){\rule[-0.350pt]{0.700pt}{0.712pt}}
\put(388,604){\rule[-0.350pt]{0.700pt}{0.712pt}}
\put(389,601){\rule[-0.350pt]{0.700pt}{0.712pt}}
\put(390,598){\rule[-0.350pt]{0.700pt}{0.712pt}}
\put(391,595){\rule[-0.350pt]{0.700pt}{0.712pt}}
\put(392,592){\rule[-0.350pt]{0.700pt}{0.712pt}}
\put(393,590){\rule[-0.350pt]{0.700pt}{0.712pt}}
\put(394,588){\usebox{\plotpoint}}
\put(395,586){\usebox{\plotpoint}}
\put(396,584){\usebox{\plotpoint}}
\put(397,582){\usebox{\plotpoint}}
\put(398,580){\usebox{\plotpoint}}
\put(399,578){\usebox{\plotpoint}}
\put(400,576){\usebox{\plotpoint}}
\put(401,574){\usebox{\plotpoint}}
\put(402,572){\usebox{\plotpoint}}
\put(403,570){\usebox{\plotpoint}}
\put(404,568){\usebox{\plotpoint}}
\put(405,566){\usebox{\plotpoint}}
\put(406,564){\usebox{\plotpoint}}
\put(407,562){\usebox{\plotpoint}}
\put(408,560){\usebox{\plotpoint}}
\put(409,558){\usebox{\plotpoint}}
\put(410,556){\usebox{\plotpoint}}
\put(411,554){\usebox{\plotpoint}}
\put(412,552){\usebox{\plotpoint}}
\put(413,550){\usebox{\plotpoint}}
\put(414,549){\usebox{\plotpoint}}
\put(415,547){\usebox{\plotpoint}}
\put(416,546){\usebox{\plotpoint}}
\put(417,544){\usebox{\plotpoint}}
\put(418,543){\usebox{\plotpoint}}
\put(419,542){\usebox{\plotpoint}}
\put(420,540){\usebox{\plotpoint}}
\put(421,539){\usebox{\plotpoint}}
\put(422,538){\usebox{\plotpoint}}
\put(423,536){\usebox{\plotpoint}}
\put(424,535){\usebox{\plotpoint}}
\put(425,533){\usebox{\plotpoint}}
\put(426,532){\usebox{\plotpoint}}
\put(427,531){\usebox{\plotpoint}}
\put(428,529){\usebox{\plotpoint}}
\put(429,528){\usebox{\plotpoint}}
\put(430,527){\usebox{\plotpoint}}
\put(431,525){\usebox{\plotpoint}}
\put(432,524){\usebox{\plotpoint}}
\put(433,523){\usebox{\plotpoint}}
\put(434,521){\usebox{\plotpoint}}
\put(435,520){\usebox{\plotpoint}}
\put(436,519){\usebox{\plotpoint}}
\put(437,519){\usebox{\plotpoint}}
\put(438,518){\usebox{\plotpoint}}
\put(439,517){\usebox{\plotpoint}}
\put(440,516){\usebox{\plotpoint}}
\put(441,515){\usebox{\plotpoint}}
\put(442,514){\usebox{\plotpoint}}
\put(443,513){\usebox{\plotpoint}}
\put(444,512){\usebox{\plotpoint}}
\put(445,511){\usebox{\plotpoint}}
\put(446,510){\usebox{\plotpoint}}
\put(448,509){\usebox{\plotpoint}}
\put(449,508){\usebox{\plotpoint}}
\put(450,507){\usebox{\plotpoint}}
\put(451,506){\usebox{\plotpoint}}
\put(452,505){\usebox{\plotpoint}}
\put(453,504){\usebox{\plotpoint}}
\put(454,503){\usebox{\plotpoint}}
\put(455,502){\usebox{\plotpoint}}
\put(456,501){\usebox{\plotpoint}}
\put(457,500){\usebox{\plotpoint}}
\put(459,499){\usebox{\plotpoint}}
\put(460,498){\usebox{\plotpoint}}
\put(461,497){\usebox{\plotpoint}}
\put(462,496){\usebox{\plotpoint}}
\put(463,495){\usebox{\plotpoint}}
\put(464,494){\usebox{\plotpoint}}
\put(465,493){\usebox{\plotpoint}}
\put(466,492){\usebox{\plotpoint}}
\put(467,491){\usebox{\plotpoint}}
\put(468,490){\usebox{\plotpoint}}
\put(470,489){\usebox{\plotpoint}}
\put(471,488){\usebox{\plotpoint}}
\put(472,487){\usebox{\plotpoint}}
\put(473,486){\usebox{\plotpoint}}
\put(474,485){\usebox{\plotpoint}}
\put(475,484){\usebox{\plotpoint}}
\put(476,483){\usebox{\plotpoint}}
\put(477,482){\usebox{\plotpoint}}
\put(478,481){\usebox{\plotpoint}}
\put(480,480){\usebox{\plotpoint}}
\put(481,479){\usebox{\plotpoint}}
\put(483,478){\usebox{\plotpoint}}
\put(485,477){\usebox{\plotpoint}}
\put(486,476){\usebox{\plotpoint}}
\put(488,475){\usebox{\plotpoint}}
\put(490,474){\usebox{\plotpoint}}
\put(491,473){\usebox{\plotpoint}}
\put(493,472){\usebox{\plotpoint}}
\put(495,471){\usebox{\plotpoint}}
\put(496,470){\usebox{\plotpoint}}
\put(498,469){\usebox{\plotpoint}}
\put(500,468){\usebox{\plotpoint}}
\put(502,467){\usebox{\plotpoint}}
\put(503,466){\usebox{\plotpoint}}
\put(505,465){\usebox{\plotpoint}}
\put(507,464){\usebox{\plotpoint}}
\put(508,463){\usebox{\plotpoint}}
\put(510,462){\usebox{\plotpoint}}
\put(512,461){\usebox{\plotpoint}}
\put(513,460){\usebox{\plotpoint}}
\put(515,459){\usebox{\plotpoint}}
\put(517,458){\usebox{\plotpoint}}
\put(518,457){\usebox{\plotpoint}}
\put(520,456){\usebox{\plotpoint}}
\put(522,455){\usebox{\plotpoint}}
\put(524,454){\usebox{\plotpoint}}
\put(526,453){\usebox{\plotpoint}}
\put(528,452){\usebox{\plotpoint}}
\put(530,451){\usebox{\plotpoint}}
\put(533,450){\usebox{\plotpoint}}
\put(535,449){\usebox{\plotpoint}}
\put(537,448){\usebox{\plotpoint}}
\put(539,447){\usebox{\plotpoint}}
\put(542,446){\usebox{\plotpoint}}
\put(544,445){\usebox{\plotpoint}}
\put(546,444){\usebox{\plotpoint}}
\put(548,443){\usebox{\plotpoint}}
\put(551,442){\usebox{\plotpoint}}
\put(553,441){\usebox{\plotpoint}}
\put(555,440){\usebox{\plotpoint}}
\put(557,439){\usebox{\plotpoint}}
\put(560,438){\usebox{\plotpoint}}
\put(562,437){\usebox{\plotpoint}}
\put(564,436){\usebox{\plotpoint}}
\end{picture}
}\parbox{1.7in}{
\setlength{\unitlength}{0.240900pt}
\ifx\plotpoint\undefined\newsavebox{\plotpoint}\fi
\sbox{\plotpoint}{\rule[-0.175pt]{0.350pt}{0.350pt}}%
\begin{picture}(674,1259)(0,0)
\tenrm
\sbox{\plotpoint}{\rule[-0.175pt]{0.350pt}{0.350pt}}%
\put(264,158){\rule[-0.175pt]{83.351pt}{0.350pt}}
\put(264,158){\rule[-0.175pt]{0.350pt}{238.009pt}}
\put(264,378){\rule[-0.175pt]{4.818pt}{0.350pt}}
\put(242,378){\makebox(0,0)[r]{$0.1$}}
\put(590,378){\rule[-0.175pt]{4.818pt}{0.350pt}}
\put(264,597){\rule[-0.175pt]{4.818pt}{0.350pt}}
\put(242,597){\makebox(0,0)[r]{$0.2$}}
\put(590,597){\rule[-0.175pt]{4.818pt}{0.350pt}}
\put(264,817){\rule[-0.175pt]{4.818pt}{0.350pt}}
\put(242,817){\makebox(0,0)[r]{$0.3$}}
\put(590,817){\rule[-0.175pt]{4.818pt}{0.350pt}}
\put(264,1036){\rule[-0.175pt]{4.818pt}{0.350pt}}
\put(242,1036){\makebox(0,0)[r]{$0.4$}}
\put(590,1036){\rule[-0.175pt]{4.818pt}{0.350pt}}
\put(351,158){\rule[-0.175pt]{0.350pt}{4.818pt}}
\put(351,113){\makebox(0,0){$1$}}
\put(351,1126){\rule[-0.175pt]{0.350pt}{4.818pt}}
\put(437,158){\rule[-0.175pt]{0.350pt}{4.818pt}}
\put(437,113){\makebox(0,0){$2$}}
\put(437,1126){\rule[-0.175pt]{0.350pt}{4.818pt}}
\put(524,158){\rule[-0.175pt]{0.350pt}{4.818pt}}
\put(524,113){\makebox(0,0){$3$}}
\put(524,1126){\rule[-0.175pt]{0.350pt}{4.818pt}}
\put(264,158){\rule[-0.175pt]{83.351pt}{0.350pt}}
\put(610,158){\rule[-0.175pt]{0.350pt}{238.009pt}}
\put(264,1146){\rule[-0.175pt]{83.351pt}{0.350pt}}
\put(437,68){\makebox(0,0){$q\,a$}}
\put(437,1064){\makebox(0,0)[l]{$\beta=6.2$}}
\put(264,158){\rule[-0.175pt]{0.350pt}{238.009pt}}
\put(559,465){\circle{24}}
\put(530,479){\circle{24}}
\put(524,481){\circle{24}}
\put(493,498){\circle{24}}
\put(485,507){\circle{24}}
\put(477,516){\circle{24}}
\put(358,615){\circle{24}}
\put(359,630){\circle{24}}
\put(348,674){\circle{24}}
\put(357,656){\circle{24}}
\put(333,733){\circle{24}}
\put(301,970){\circle{24}}
\sbox{\plotpoint}{\rule[-0.350pt]{0.700pt}{0.700pt}}%
\put(299,1089){\usebox{\plotpoint}}
\put(299,1067){\rule[-0.350pt]{0.700pt}{5.149pt}}
\put(300,1046){\rule[-0.350pt]{0.700pt}{5.149pt}}
\put(301,1024){\rule[-0.350pt]{0.700pt}{5.149pt}}
\put(302,1003){\rule[-0.350pt]{0.700pt}{5.149pt}}
\put(303,982){\rule[-0.350pt]{0.700pt}{5.149pt}}
\put(304,960){\rule[-0.350pt]{0.700pt}{5.149pt}}
\put(305,939){\rule[-0.350pt]{0.700pt}{5.149pt}}
\put(306,918){\rule[-0.350pt]{0.700pt}{5.149pt}}
\put(307,907){\rule[-0.350pt]{0.700pt}{2.596pt}}
\put(308,896){\rule[-0.350pt]{0.700pt}{2.596pt}}
\put(309,885){\rule[-0.350pt]{0.700pt}{2.596pt}}
\put(310,874){\rule[-0.350pt]{0.700pt}{2.596pt}}
\put(311,864){\rule[-0.350pt]{0.700pt}{2.596pt}}
\put(312,853){\rule[-0.350pt]{0.700pt}{2.596pt}}
\put(313,842){\rule[-0.350pt]{0.700pt}{2.596pt}}
\put(314,831){\rule[-0.350pt]{0.700pt}{2.596pt}}
\put(315,821){\rule[-0.350pt]{0.700pt}{2.596pt}}
\put(316,814){\rule[-0.350pt]{0.700pt}{1.633pt}}
\put(317,807){\rule[-0.350pt]{0.700pt}{1.633pt}}
\put(318,800){\rule[-0.350pt]{0.700pt}{1.633pt}}
\put(319,793){\rule[-0.350pt]{0.700pt}{1.633pt}}
\put(320,787){\rule[-0.350pt]{0.700pt}{1.633pt}}
\put(321,780){\rule[-0.350pt]{0.700pt}{1.633pt}}
\put(322,773){\rule[-0.350pt]{0.700pt}{1.633pt}}
\put(323,766){\rule[-0.350pt]{0.700pt}{1.633pt}}
\put(324,760){\rule[-0.350pt]{0.700pt}{1.633pt}}
\put(325,754){\rule[-0.350pt]{0.700pt}{1.415pt}}
\put(326,748){\rule[-0.350pt]{0.700pt}{1.415pt}}
\put(327,742){\rule[-0.350pt]{0.700pt}{1.415pt}}
\put(328,736){\rule[-0.350pt]{0.700pt}{1.415pt}}
\put(329,730){\rule[-0.350pt]{0.700pt}{1.415pt}}
\put(330,724){\rule[-0.350pt]{0.700pt}{1.415pt}}
\put(331,718){\rule[-0.350pt]{0.700pt}{1.415pt}}
\put(332,713){\rule[-0.350pt]{0.700pt}{1.415pt}}
\put(333,709){\rule[-0.350pt]{0.700pt}{0.857pt}}
\put(334,705){\rule[-0.350pt]{0.700pt}{0.857pt}}
\put(335,702){\rule[-0.350pt]{0.700pt}{0.857pt}}
\put(336,698){\rule[-0.350pt]{0.700pt}{0.857pt}}
\put(337,695){\rule[-0.350pt]{0.700pt}{0.857pt}}
\put(338,691){\rule[-0.350pt]{0.700pt}{0.857pt}}
\put(339,688){\rule[-0.350pt]{0.700pt}{0.857pt}}
\put(340,684){\rule[-0.350pt]{0.700pt}{0.857pt}}
\put(341,681){\rule[-0.350pt]{0.700pt}{0.857pt}}
\put(342,678){\rule[-0.350pt]{0.700pt}{0.723pt}}
\put(343,675){\rule[-0.350pt]{0.700pt}{0.723pt}}
\put(344,672){\rule[-0.350pt]{0.700pt}{0.723pt}}
\put(345,669){\rule[-0.350pt]{0.700pt}{0.723pt}}
\put(346,666){\rule[-0.350pt]{0.700pt}{0.723pt}}
\put(347,663){\rule[-0.350pt]{0.700pt}{0.723pt}}
\put(348,660){\rule[-0.350pt]{0.700pt}{0.723pt}}
\put(349,657){\rule[-0.350pt]{0.700pt}{0.723pt}}
\put(350,654){\rule[-0.350pt]{0.700pt}{0.723pt}}
\put(351,651){\usebox{\plotpoint}}
\put(352,649){\usebox{\plotpoint}}
\put(353,647){\usebox{\plotpoint}}
\put(354,644){\usebox{\plotpoint}}
\put(355,642){\usebox{\plotpoint}}
\put(356,640){\usebox{\plotpoint}}
\put(357,638){\usebox{\plotpoint}}
\put(358,635){\usebox{\plotpoint}}
\put(359,633){\usebox{\plotpoint}}
\put(360,631){\usebox{\plotpoint}}
\put(361,628){\usebox{\plotpoint}}
\put(362,626){\usebox{\plotpoint}}
\put(363,624){\usebox{\plotpoint}}
\put(364,622){\usebox{\plotpoint}}
\put(365,619){\usebox{\plotpoint}}
\put(366,617){\usebox{\plotpoint}}
\put(367,615){\usebox{\plotpoint}}
\put(368,612){\usebox{\plotpoint}}
\put(369,610){\usebox{\plotpoint}}
\put(370,608){\usebox{\plotpoint}}
\put(371,606){\usebox{\plotpoint}}
\put(372,604){\usebox{\plotpoint}}
\put(373,603){\usebox{\plotpoint}}
\put(374,601){\usebox{\plotpoint}}
\put(375,600){\usebox{\plotpoint}}
\put(376,598){\usebox{\plotpoint}}
\put(377,597){\usebox{\plotpoint}}
\put(378,595){\usebox{\plotpoint}}
\put(379,594){\usebox{\plotpoint}}
\put(380,592){\usebox{\plotpoint}}
\put(381,591){\usebox{\plotpoint}}
\put(382,589){\usebox{\plotpoint}}
\put(383,588){\usebox{\plotpoint}}
\put(384,586){\usebox{\plotpoint}}
\put(385,585){\usebox{\plotpoint}}
\put(386,583){\usebox{\plotpoint}}
\put(387,582){\usebox{\plotpoint}}
\put(388,580){\usebox{\plotpoint}}
\put(389,579){\usebox{\plotpoint}}
\put(390,577){\usebox{\plotpoint}}
\put(391,576){\usebox{\plotpoint}}
\put(392,574){\usebox{\plotpoint}}
\put(393,573){\usebox{\plotpoint}}
\put(394,571){\usebox{\plotpoint}}
\put(395,570){\usebox{\plotpoint}}
\put(396,569){\usebox{\plotpoint}}
\put(397,568){\usebox{\plotpoint}}
\put(398,567){\usebox{\plotpoint}}
\put(399,566){\usebox{\plotpoint}}
\put(400,565){\usebox{\plotpoint}}
\put(401,564){\usebox{\plotpoint}}
\put(402,563){\usebox{\plotpoint}}
\put(403,562){\usebox{\plotpoint}}
\put(404,561){\usebox{\plotpoint}}
\put(405,560){\usebox{\plotpoint}}
\put(406,559){\usebox{\plotpoint}}
\put(407,558){\usebox{\plotpoint}}
\put(408,557){\usebox{\plotpoint}}
\put(409,556){\usebox{\plotpoint}}
\put(410,555){\usebox{\plotpoint}}
\put(411,554){\usebox{\plotpoint}}
\put(412,553){\usebox{\plotpoint}}
\put(413,552){\usebox{\plotpoint}}
\put(414,551){\usebox{\plotpoint}}
\put(415,551){\usebox{\plotpoint}}
\put(415,551){\usebox{\plotpoint}}
\put(416,550){\usebox{\plotpoint}}
\put(417,549){\usebox{\plotpoint}}
\put(418,548){\usebox{\plotpoint}}
\put(419,547){\usebox{\plotpoint}}
\put(420,546){\usebox{\plotpoint}}
\put(421,545){\usebox{\plotpoint}}
\put(422,544){\usebox{\plotpoint}}
\put(423,543){\usebox{\plotpoint}}
\put(424,542){\usebox{\plotpoint}}
\put(426,541){\usebox{\plotpoint}}
\put(427,540){\usebox{\plotpoint}}
\put(428,539){\usebox{\plotpoint}}
\put(429,538){\usebox{\plotpoint}}
\put(430,537){\usebox{\plotpoint}}
\put(431,536){\usebox{\plotpoint}}
\put(432,535){\usebox{\plotpoint}}
\put(433,534){\usebox{\plotpoint}}
\put(434,533){\usebox{\plotpoint}}
\put(435,532){\usebox{\plotpoint}}
\put(437,531){\usebox{\plotpoint}}
\put(438,530){\usebox{\plotpoint}}
\put(440,529){\usebox{\plotpoint}}
\put(441,528){\usebox{\plotpoint}}
\put(443,527){\usebox{\plotpoint}}
\put(445,526){\usebox{\plotpoint}}
\put(446,525){\usebox{\plotpoint}}
\put(448,524){\usebox{\plotpoint}}
\put(450,523){\usebox{\plotpoint}}
\put(451,522){\usebox{\plotpoint}}
\put(453,521){\usebox{\plotpoint}}
\put(455,520){\usebox{\plotpoint}}
\put(456,519){\usebox{\plotpoint}}
\put(458,518){\usebox{\plotpoint}}
\put(460,517){\usebox{\plotpoint}}
\put(461,516){\usebox{\plotpoint}}
\put(463,515){\usebox{\plotpoint}}
\put(465,514){\usebox{\plotpoint}}
\put(466,513){\usebox{\plotpoint}}
\put(468,512){\usebox{\plotpoint}}
\put(470,511){\usebox{\plotpoint}}
\put(471,510){\usebox{\plotpoint}}
\put(473,509){\usebox{\plotpoint}}
\put(475,508){\usebox{\plotpoint}}
\put(476,507){\usebox{\plotpoint}}
\put(478,506){\usebox{\plotpoint}}
\put(479,505){\usebox{\plotpoint}}
\put(482,504){\usebox{\plotpoint}}
\put(484,503){\usebox{\plotpoint}}
\put(487,502){\usebox{\plotpoint}}
\put(489,501){\usebox{\plotpoint}}
\put(492,500){\usebox{\plotpoint}}
\put(494,499){\usebox{\plotpoint}}
\put(497,498){\usebox{\plotpoint}}
\put(499,497){\usebox{\plotpoint}}
\put(502,496){\usebox{\plotpoint}}
\put(504,495){\usebox{\plotpoint}}
\put(506,494){\usebox{\plotpoint}}
\put(509,493){\usebox{\plotpoint}}
\put(511,492){\usebox{\plotpoint}}
\put(514,491){\usebox{\plotpoint}}
\put(516,490){\usebox{\plotpoint}}
\put(519,489){\usebox{\plotpoint}}
\put(521,488){\usebox{\plotpoint}}
\put(524,487){\usebox{\plotpoint}}
\put(526,486){\usebox{\plotpoint}}
\put(529,485){\usebox{\plotpoint}}
\put(532,484){\usebox{\plotpoint}}
\put(535,483){\usebox{\plotpoint}}
\put(538,482){\usebox{\plotpoint}}
\put(541,481){\usebox{\plotpoint}}
\put(544,480){\usebox{\plotpoint}}
\put(546,479){\usebox{\plotpoint}}
\put(549,478){\usebox{\plotpoint}}
\put(552,477){\usebox{\plotpoint}}
\put(555,476){\usebox{\plotpoint}}
\put(558,475){\usebox{\plotpoint}}
\put(561,474){\usebox{\plotpoint}}
\put(564,473){\usebox{\plotpoint}}
\put(566,472){\usebox{\plotpoint}}
\end{picture}
}\parbox{1.7in}{
\setlength{\unitlength}{0.240900pt}
\ifx\plotpoint\undefined\newsavebox{\plotpoint}\fi
\sbox{\plotpoint}{\rule[-0.175pt]{0.350pt}{0.350pt}}%
\begin{picture}(674,1259)(0,0)
\tenrm
\sbox{\plotpoint}{\rule[-0.175pt]{0.350pt}{0.350pt}}%
\put(264,158){\rule[-0.175pt]{83.351pt}{0.350pt}}
\put(264,158){\rule[-0.175pt]{0.350pt}{238.009pt}}
\put(264,570){\rule[-0.175pt]{4.818pt}{0.350pt}}
\put(242,570){\makebox(0,0)[r]{$0.05$}}
\put(590,570){\rule[-0.175pt]{4.818pt}{0.350pt}}
\put(264,981){\rule[-0.175pt]{4.818pt}{0.350pt}}
\put(242,981){\makebox(0,0)[r]{$0.1$}}
\put(590,981){\rule[-0.175pt]{4.818pt}{0.350pt}}
\put(351,158){\rule[-0.175pt]{0.350pt}{4.818pt}}
\put(351,113){\makebox(0,0){$1$}}
\put(351,1126){\rule[-0.175pt]{0.350pt}{4.818pt}}
\put(437,158){\rule[-0.175pt]{0.350pt}{4.818pt}}
\put(437,113){\makebox(0,0){$2$}}
\put(437,1126){\rule[-0.175pt]{0.350pt}{4.818pt}}
\put(524,158){\rule[-0.175pt]{0.350pt}{4.818pt}}
\put(524,113){\makebox(0,0){$3$}}
\put(524,1126){\rule[-0.175pt]{0.350pt}{4.818pt}}
\put(264,158){\rule[-0.175pt]{83.351pt}{0.350pt}}
\put(610,158){\rule[-0.175pt]{0.350pt}{238.009pt}}
\put(264,1146){\rule[-0.175pt]{83.351pt}{0.350pt}}
\put(437,68){\makebox(0,0){$q\,a$}}
\put(437,1064){\makebox(0,0)[l]{$\beta=9$}}
\put(264,158){\rule[-0.175pt]{0.350pt}{238.009pt}}
\put(559,759){\circle{24}}
\put(530,767){\circle{24}}
\put(524,776){\circle{24}}
\put(493,784){\circle{24}}
\put(485,792){\circle{24}}
\put(477,792){\circle{24}}
\put(358,874){\circle{24}}
\put(359,874){\circle{24}}
\put(348,891){\circle{24}}
\put(357,883){\circle{24}}
\put(333,915){\circle{24}}
\put(301,990){\circle{24}}
\sbox{\plotpoint}{\rule[-0.350pt]{0.700pt}{0.700pt}}%
\put(299,1022){\usebox{\plotpoint}}
\put(299,1016){\rule[-0.350pt]{0.700pt}{1.235pt}}
\put(300,1011){\rule[-0.350pt]{0.700pt}{1.235pt}}
\put(301,1006){\rule[-0.350pt]{0.700pt}{1.235pt}}
\put(302,1001){\rule[-0.350pt]{0.700pt}{1.235pt}}
\put(303,996){\rule[-0.350pt]{0.700pt}{1.235pt}}
\put(304,991){\rule[-0.350pt]{0.700pt}{1.235pt}}
\put(305,986){\rule[-0.350pt]{0.700pt}{1.235pt}}
\put(306,981){\rule[-0.350pt]{0.700pt}{1.235pt}}
\put(307,978){\usebox{\plotpoint}}
\put(308,975){\usebox{\plotpoint}}
\put(309,972){\usebox{\plotpoint}}
\put(310,970){\usebox{\plotpoint}}
\put(311,967){\usebox{\plotpoint}}
\put(312,964){\usebox{\plotpoint}}
\put(313,962){\usebox{\plotpoint}}
\put(314,959){\usebox{\plotpoint}}
\put(315,957){\usebox{\plotpoint}}
\put(316,954){\usebox{\plotpoint}}
\put(317,951){\usebox{\plotpoint}}
\put(318,948){\usebox{\plotpoint}}
\put(319,945){\usebox{\plotpoint}}
\put(320,943){\usebox{\plotpoint}}
\put(321,940){\usebox{\plotpoint}}
\put(322,937){\usebox{\plotpoint}}
\put(323,934){\usebox{\plotpoint}}
\put(324,932){\usebox{\plotpoint}}
\put(325,929){\usebox{\plotpoint}}
\put(326,927){\usebox{\plotpoint}}
\put(327,925){\usebox{\plotpoint}}
\put(328,923){\usebox{\plotpoint}}
\put(329,921){\usebox{\plotpoint}}
\put(330,919){\usebox{\plotpoint}}
\put(331,917){\usebox{\plotpoint}}
\put(332,915){\usebox{\plotpoint}}
\put(333,913){\usebox{\plotpoint}}
\put(334,911){\usebox{\plotpoint}}
\put(335,909){\usebox{\plotpoint}}
\put(336,907){\usebox{\plotpoint}}
\put(337,906){\usebox{\plotpoint}}
\put(338,904){\usebox{\plotpoint}}
\put(339,902){\usebox{\plotpoint}}
\put(340,900){\usebox{\plotpoint}}
\put(341,899){\usebox{\plotpoint}}
\put(342,897){\usebox{\plotpoint}}
\put(343,895){\usebox{\plotpoint}}
\put(344,893){\usebox{\plotpoint}}
\put(345,891){\usebox{\plotpoint}}
\put(346,890){\usebox{\plotpoint}}
\put(347,888){\usebox{\plotpoint}}
\put(348,886){\usebox{\plotpoint}}
\put(349,884){\usebox{\plotpoint}}
\put(350,883){\usebox{\plotpoint}}
\put(351,881){\usebox{\plotpoint}}
\put(352,880){\usebox{\plotpoint}}
\put(353,879){\usebox{\plotpoint}}
\put(354,878){\usebox{\plotpoint}}
\put(355,877){\usebox{\plotpoint}}
\put(356,875){\usebox{\plotpoint}}
\put(357,874){\usebox{\plotpoint}}
\put(358,873){\usebox{\plotpoint}}
\put(359,872){\usebox{\plotpoint}}
\put(360,871){\usebox{\plotpoint}}
\put(361,869){\usebox{\plotpoint}}
\put(362,868){\usebox{\plotpoint}}
\put(363,867){\usebox{\plotpoint}}
\put(364,866){\usebox{\plotpoint}}
\put(365,865){\usebox{\plotpoint}}
\put(366,863){\usebox{\plotpoint}}
\put(367,862){\usebox{\plotpoint}}
\put(368,861){\usebox{\plotpoint}}
\put(369,860){\usebox{\plotpoint}}
\put(370,859){\usebox{\plotpoint}}
\put(371,858){\usebox{\plotpoint}}
\put(372,858){\usebox{\plotpoint}}
\put(373,857){\usebox{\plotpoint}}
\put(374,856){\usebox{\plotpoint}}
\put(375,855){\usebox{\plotpoint}}
\put(377,854){\usebox{\plotpoint}}
\put(378,853){\usebox{\plotpoint}}
\put(379,852){\usebox{\plotpoint}}
\put(381,851){\usebox{\plotpoint}}
\put(382,850){\usebox{\plotpoint}}
\put(383,849){\usebox{\plotpoint}}
\put(384,848){\usebox{\plotpoint}}
\put(386,847){\usebox{\plotpoint}}
\put(387,846){\usebox{\plotpoint}}
\put(388,845){\usebox{\plotpoint}}
\put(390,844){\usebox{\plotpoint}}
\put(391,843){\usebox{\plotpoint}}
\put(392,842){\usebox{\plotpoint}}
\put(394,841){\usebox{\plotpoint}}
\put(395,840){\usebox{\plotpoint}}
\put(396,839){\usebox{\plotpoint}}
\put(397,838){\usebox{\plotpoint}}
\put(399,837){\usebox{\plotpoint}}
\put(400,836){\usebox{\plotpoint}}
\put(401,835){\usebox{\plotpoint}}
\put(403,834){\usebox{\plotpoint}}
\put(404,833){\usebox{\plotpoint}}
\put(405,832){\usebox{\plotpoint}}
\put(407,831){\usebox{\plotpoint}}
\put(408,830){\usebox{\plotpoint}}
\put(409,829){\usebox{\plotpoint}}
\put(411,828){\usebox{\plotpoint}}
\put(412,827){\usebox{\plotpoint}}
\put(413,826){\usebox{\plotpoint}}
\put(415,825){\usebox{\plotpoint}}
\put(416,824){\usebox{\plotpoint}}
\put(417,823){\usebox{\plotpoint}}
\put(418,822){\usebox{\plotpoint}}
\put(420,821){\usebox{\plotpoint}}
\put(421,820){\usebox{\plotpoint}}
\put(422,819){\usebox{\plotpoint}}
\put(424,818){\usebox{\plotpoint}}
\put(425,817){\usebox{\plotpoint}}
\put(426,816){\usebox{\plotpoint}}
\put(427,815){\usebox{\plotpoint}}
\put(429,814){\usebox{\plotpoint}}
\put(430,813){\usebox{\plotpoint}}
\put(431,812){\usebox{\plotpoint}}
\put(433,811){\usebox{\plotpoint}}
\put(434,810){\usebox{\plotpoint}}
\put(435,809){\usebox{\plotpoint}}
\put(437,808){\usebox{\plotpoint}}
\put(439,807){\usebox{\plotpoint}}
\put(442,806){\usebox{\plotpoint}}
\put(445,805){\usebox{\plotpoint}}
\put(447,804){\usebox{\plotpoint}}
\put(450,803){\usebox{\plotpoint}}
\put(453,802){\usebox{\plotpoint}}
\put(455,801){\usebox{\plotpoint}}
\put(458,800){\usebox{\plotpoint}}
\put(461,799){\usebox{\plotpoint}}
\put(463,798){\usebox{\plotpoint}}
\put(466,797){\usebox{\plotpoint}}
\put(469,796){\usebox{\plotpoint}}
\put(471,795){\usebox{\plotpoint}}
\put(474,794){\usebox{\plotpoint}}
\put(477,793){\usebox{\plotpoint}}
\put(480,792){\usebox{\plotpoint}}
\put(482,791){\usebox{\plotpoint}}
\put(485,790){\usebox{\plotpoint}}
\put(488,789){\usebox{\plotpoint}}
\put(491,788){\usebox{\plotpoint}}
\put(493,787){\usebox{\plotpoint}}
\put(496,786){\usebox{\plotpoint}}
\put(499,785){\usebox{\plotpoint}}
\put(502,784){\usebox{\plotpoint}}
\put(504,783){\usebox{\plotpoint}}
\put(507,782){\usebox{\plotpoint}}
\put(510,781){\usebox{\plotpoint}}
\put(513,780){\usebox{\plotpoint}}
\put(515,779){\usebox{\plotpoint}}
\put(518,778){\usebox{\plotpoint}}
\put(521,777){\usebox{\plotpoint}}
\put(524,776){\usebox{\plotpoint}}
\put(526,775){\usebox{\plotpoint}}
\put(529,774){\usebox{\plotpoint}}
\put(531,773){\usebox{\plotpoint}}
\put(534,772){\usebox{\plotpoint}}
\put(536,771){\usebox{\plotpoint}}
\put(539,770){\usebox{\plotpoint}}
\put(541,769){\usebox{\plotpoint}}
\put(544,768){\usebox{\plotpoint}}
\put(546,767){\usebox{\plotpoint}}
\put(549,766){\usebox{\plotpoint}}
\put(551,765){\usebox{\plotpoint}}
\put(554,764){\usebox{\plotpoint}}
\put(556,763){\usebox{\plotpoint}}
\put(559,762){\usebox{\plotpoint}}
\put(561,761){\usebox{\plotpoint}}
\put(564,760){\usebox{\plotpoint}}
\end{picture}
}
\caption{Values of $\alpha_V(q)$ for a range of~$q$'s as determined from
lattice QCD measurements at various $\beta$'s. The data points (circles) are
measured values (with negligible statistical errors) obtained by fitting
second-order perturbation theory to Monte Carlo simulation data for various
short-distance quantities.  The solid line shows the variation in~$\alpha_V(q)$
expected from two-loop perturbation theory.}
 \label{alpha-evol}
\end{figure}
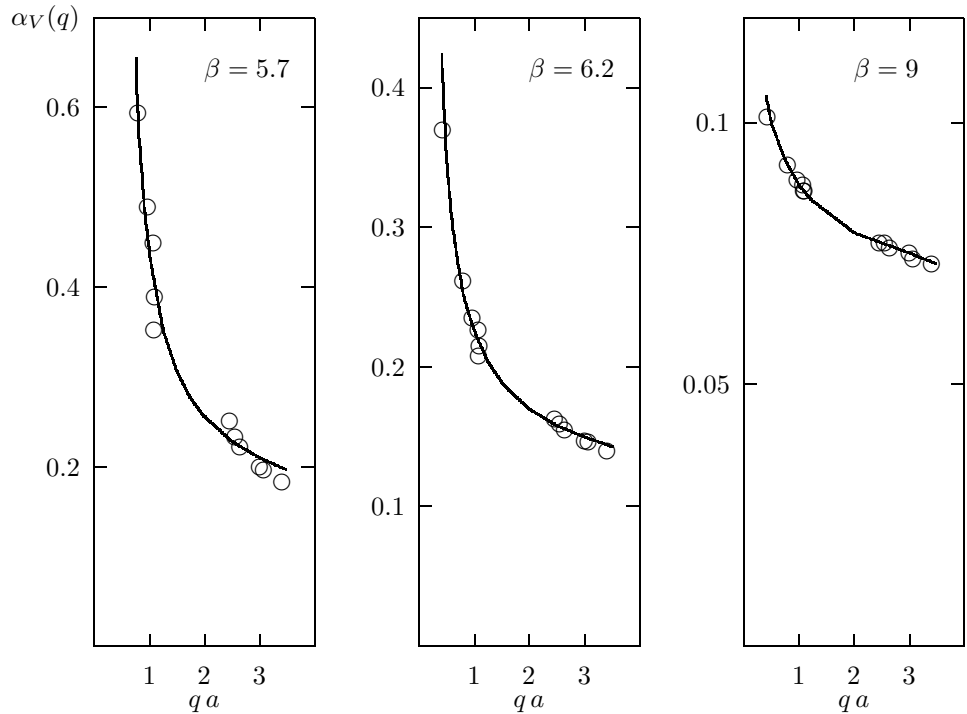

The $q$-dependence of~$\alpha_V(q)$ can be determined directly from
simulations. The data in Figure~\ref{alpha-evol} are obtained by
fitting second-order expansions in~$\alpha_V(q^*)$ to simulation results for
the six smallest Creutz ratios and the six smallest Wilson loops. The value
of~$\alpha_V(q^*)$ obtained from each fit is plotted versus the~$q^*$ for
that quantity. The~$q^*$'s for the twelve quantities used here range
from~$0.4/a$ (for~$\chi_{44}$) to~$3.4/a$ (for $-\ln W_{11}$). These
simulation results may be compared with the two-loop perturbative
prediction for~$\alpha_V(q)$ (solid line), arbitrarily
normalized so that the curve passes through the point for~$-\ln
W_{22}$. The Monte Carlo data are quite consistent with perturbation
theory, even at $\beta = 5.7$. The smallest~$q^*$'s shown are on the order
of several hundred~MeV.

These data, and much more (see later sections), show that lattice
perturbation theory for QCD is very effective even at momenta as low
as~1~GeV.
Indeed it is about as effective as continuum perturbation theory.
Perturbation theory seems to work so long as $\alpha_V(q)<1$. Possibly the
most important consequence of this result is that it becomes likely that
perturbative improvement of the QCD lagrangian will work even
with lattice spacings as large as $a = 1/2~{\rm fm}$.
However, there remains one problem:  perturbative
expansions for the couplings in the corrected lagrangian tend to have very
large coefficients. The origin of this problem and its cure are the subject
of the next section.

\section{Tadpole Improvement}
The very large renormalization factor relating the bare lattice
coupling~$\alpha_\lat$ to the continuum coupling~$\alpha_V$ is one of many
examples where a large renormalization is required to relate a lattice
quantity to its continuum analogue. These large renormalizations all have a
common source: the compact nature of the link variable used in lattice
theories.

We design lattice operators by mapping them onto analogous operators in the
continuum theory. For gauge fields, the mapping is based upon the expansion
\be
\U\mu(x) \equiv \exp(-iga\overline{A}_\mu) \to 1 - iga\A\mu(x).
\ee
This mapping seems plausible when~$a$ is small, but it is misleading
in the quantum theory since further corrections do not vanish as powers
of~$a$. Instead one gets terms like $\half(iga\A\mu)^2$ whose vacuum
expectation value is suppressed only by $g^2$\,---\,the explicit~$a^2$
cancels a $1/a^2$ from the ultraviolet divergent $\avg{\A\mu^2}$.
These corrections turn out to be uncomfortably large. For example,
\be \label{trU}
\avg{\third\Tr\U\mu}_{LG} \approx 0.82
\ee
at $\beta=5.7$, while
$\avg{\third\Tr(1-iga\A\mu)} = 1$. These very divergent loop contributions
are referred to as ``tadpole'' contributions.

The tadpoles spoil our intuition about the connection between lattice
operators and the continuum, leading to unexpected renormalizations. To
regain this intuition we need to sharpen the relation between~$\U\mu$ and
the continuum gauge field~$\A\mu$. Results like that in \eq{trU} suggest
that the appropriate connection with the continuum is
\be
\U\mu(x) \to u_0\,(1-iga\A\mu(x))
\ee
where $u_0$ can be thought of as the mean value of the link operator (gauge
invariance requires that it be an overall factor). This suggests that
lattice operators will be more continuum-like if we replace every link
operator~$\U\mu$ by a ``tadpole improved'' operator~$\Ut\mu$:
\be
\U\mu \to \Ut\mu \equiv \frac{\U\mu}{u_0}.
\ee

The precise definition of~$u_0$ is somewhat arbitrary. One could use
$\avg{\third\Tr\U\mu}_{LG}$, but it is more convenient (and
not very different) to define $u_0$ in terms of the expectation value of the
plaquette:
\be
u_0 \equiv \avg{\third\Tr\Uplaq\mu\nu}^{1/4}.
\ee
Plaquette values for several~$\beta$'s are given in
Table~\ref{alphas}. At $\beta=5.7$, $u_0$ is~0.86, making $\Ut\mu$
about 15\% smaller than $\U\mu$. This difference is considerable,
particularly in operators that involve products of several link operators.

The tadpole-improved operators~$\Ut\mu$ are much closer in their behavior to
their continuum analogues than the unimproved operators; large
tadpole renormalizations are canceled by the~$u_0$. To see what effect
tadpole improvement has on the design of lattice operators we now examine
two applications.

\subsection{The gluon action}
Our new prescription for building continuum-like operators on the lattice
suggests that
\be
\tilde{S}_{\rm gluon} = \sum \frac{6}{\tilde g_\lat^2 u_0^4} {\rm
Re}\left( \third\Tr\Uplaq\mu\nu\right) ,
\ee
with four powers of $u_0$ to correct
for the four link operators, is a better action for lattice QCD. In
particular, perturbation theory in~$\tilde\alpha_\lat\equiv\tilde
g_\lat^2/4\pi$ should be more like continuum perturbation theory (ie, no
tadpoles). Of course, this action becomes the same as the standard action if
we identify
\be
\alpha_\lat = u_0^4\,\tilde\alpha_\lat.
\ee
This relation is important because it explains why perturbative expansions
in the bare coupling~$\alpha_\lat$ fail so badly. Our tadpole analysis
indicates that $\tilde\alpha_\lat$, not $\alpha_\lat$, is the correct
expansion parameter for the theory. The difference is significant: for
example, $\tilde\alpha_\lat\approx1.8\,\alpha_\lat$ at
$\beta=5.7$. Using $\alpha_\lat$ almost guarantees that perturbation theory
will give results that are much too small.\footnote{An exception is the
expectation value of the plaquette itself (without a logarithm). Because
this operator is identical to those in the lagrangian, tadpole corrections
cancel in its expectation value. Thus the tadpole improved expansion and the
expansion in terms of~$\alpha_\lat$ are the same in lowest order, and naive,
unrenormalized lattice perturbation theory gives pretty good results for
this quantity. This is very much a special case.}

If this analysis is correct, the tadpole-improved bare
coupling~$\tilde\alpha_\lat$ should be roughly equal to the continuum
coupling~$\alpha_V(\pi/a)$. We can check this in perturbation theory which
gives a relation between $\alpha_V$ and the usual bare coupling:
\be
\alpha_V(\pi/a) = \alpha_\lat \left\{ 1 + 4.7 \alpha_V +\order(\alpha_V^2)
\right\}.
\ee
We tadpole improve this expansion by dividing out the perturbative
expansion for $1/u_0^4 = 1/\avg{\third\Tr\Uplaq\mu\nu}$ (\eq{ptplaq})
on the left-hand side:
\begin{eqnarray}
\alpha_V(\pi/a) &=& \frac{\alpha_\lat}{\avg{\third\Tr\Uplaq\mu\nu}}\,
\left\{1 + 0.513\alpha_V + \order(\alpha_V^2)\right\} \\
&=& \tilde\alpha_\lat\,\left\{1 + 0.513\alpha_V + \order(\alpha_V^2)\right\}.
\end{eqnarray}
This equation implies that $\alpha_V(\pi/a)$ and
$\tilde\alpha_\lat$ agree to within~10\%  at~$\beta=5.7$,
while~$\alpha_\lat$ is almost a factor of two smaller. Thus perturbation
theory confirms that almost all of the large renormalization factor
relating~$\alpha_V$ to~$\alpha_\lat$ is due to tadpoles.

Tadpole improvement is very important when improving lattice actions, since
the correction terms tend to have lots of link operators. If you completed
the Exercises in the earlier sections, you found that the order~$a^2$ errors
in the standard lattice action for classical QCD are removed  by replacing
\be
\Uplaq\mu\nu \to \Uplaq\mu\nu -
\frac{1}{20}\left(\Urect\mu\nu+\Urect\nu\mu\right) \qquad\mbox{(classical)},
\ee
where $\Urect\mu\nu$ is the $1\times2$~planar Wilson loop.
Since~$\Urect\mu\nu$ has two more link operators than~$\Uplaq\mu\nu$, the
tadpole improved correction in the quantum theory is given by
\be
\Uplaq\mu\nu \to \Uplaq\mu\nu -
\frac{1}{20\,u_0^2}\left(\Urect\mu\nu+\Urect\nu\mu\right)
\qquad\mbox{(tadpole-improved)}.
\ee
The extra factor of $1/u_0^2$ is important, particularly at the large
lattice spacings we want to use; without it the correction is
too small and order~$a^2$ errors are only partially removed.

\subsection{The quark action}
As a second example, consider the tadpole-improved version of Wilson's
action for heavy quarks:
\be
\tilde S_q = \sum_x \overline\psi\psi - \tilde\kappa \sum_{x,\mu}
\left\{
\overline\psi(x+a\hat\mu)\left((1+\gamma_\mu)\,\frac{\U\mu}{u_0}\right)
\psi(x) + {\rm h.c.} \right\}.
\ee
Again, this action is identical to the usual one if we relate the modified
parameters, here the ``hopping parameter''~$\tilde\kappa$, to the usual
ones by rescaling with~$u_0$:
\be
\tilde\kappa = \kappa\,u_0.
\ee

\begin{exercise}
Setting $\U\mu/u_0 = 1$, show that the quark mass in the free-quark theory
is given by $m a = 1/2\tilde\kappa - 4$.
\end{exercise}

The modified hopping parameter should be more continuum-like than the
usual one; for example, the free-quark value that gives massless quarks,
$\tilde\kappa_c = 1/8$, should be roughly correct for interacting quarks as
well. Thus a nonperturbative formula for the critical value of the
conventional hopping parameter is
\be
\kappa_c \approx 1/8u_0 ,
\ee
which implies that the critical value of the bare quark mass is
\be
m_c\,a \equiv 1/2\kappa_c -4 \approx 4(u_0-1) .
\ee
Again we turn to perturbation theory to check the validity of this
result. In perturbation theory, $m_c$ is given by
\be \label{ptkappa}
m_c\,a = -5.457 \alpha_V(2.6/a) + \order(\alpha_V^2)
\ee
where the scale of~$\alpha_V$ is determined as in
Section~\ref{definingalpha}. Pulling out a~$4(u_0-1)$ for the tadpole
contribution, this becomes
\be \label{tiptkappa}
 m_c\,a = 4\left( \avg{\third\Tr \Uplaq\mu\nu }^{1/4} -1 \right)
     - 1.268\,\alpha_V(1.0/a)  + \order(\alpha_V^2).
\ee
Thus perturbation theory confirms that the bulk of the renormalization
of~$m_c$ is due to tadpoles. Notice also that the scale in~$\alpha_V$
is smaller with tadpole improvement; this is because the very ultraviolet
tadpole contributions have been removed.

By removing the tadpole contributions,
we make the perturbation theory more convergent, and therefore more
accurate. The value of $u_0$ is obtained directly from the simulation (from
the measured plaquette expectation value), and so we need never deal with
tadpole effects in perturbation theory. The accuracy of this procedure is
illustrated in Table~\ref{kappac} where we list values for
$m_c\,a$ at different~$\beta$'s. We give results from the ordinary
perturbative expansion~(\eq{ptkappa}), from the tadpole-improved
expansion~(\eq{tiptkappa}), and from nonperturbative Monte Carlo simulations.
The tadpole-improved results agree very well with the
simulation results, and are about as accurate.
\begin{table}
\begin{center}
 \begin{tabular}{c|ccc}
$\beta$ &  perturbative & tadpole-improved & simulation
\\ \hline
5.7 &  -1.11 & -1.00 & -1.04(2) \\
6.0 &  -0.91 & -0.80 & -0.80(2) \\
6.1 &  -0.86 & -0.76 & -0.78(2) \\
6.3 &  -0.79 & -0.70 & -0.70(2) \\
 \end{tabular}
\end{center}
\caption{Values for the critical mass~$m_c\,a$ in Wilson's quark action as
computed using ordinary perturbation theory, tadpole-improved perturbation
theory, and Monte Carlo simulation.}
\label{kappac}
\end{table}

Because there are link operators in the kinetic part of the quark action,
tadpole improvement of the links affects the relation between the lattice
quark field and the continuum quark field. In the continuum limit, the
tadpole-improved lagrangian for massless quarks becomes
\be
2\tilde\kappa_c\overline\psi\gamma_\mu\partial^\mu\psi + \order(a).
\ee
This indicates that the tadpole-improved lattice quark operator is
\be
\tilde\psi = \sqrt{2\tilde\kappa_c} \psi = \psi/2 \qquad \mbox{(massless
quarks)},
\ee
where we use the fact that $\tilde\kappa_c \approx 1/8$.
This lattice operator has roughly the same normalization as the continuum
field; in particular, there are no large tadpole contributions
contributions to the renormalization constant relating them. This is
important in designing new lattice operators involving quark fields. For
example, if one wants to calculate matrix elements of the continuum current
$\overline\psi\gamma^\mu\gamma^5\psi$, then one should simulate with the
lattice operator
\be
\overline{\tilde\psi}\gamma^\mu\gamma^5\tilde\psi =
\mbox{$\frac{1}{4}$}\overline\psi\gamma^\mu\gamma^5\psi .
\ee

\begin{exercise}
Compute the tree-level correction needed to remove order~$a$ errors from
the Wilson action. Tadpole-improve your result.
\end{exercise}

\section{Nonrelativistic QCD}
As an illustration of the ideas developed in these lectures we end with a
review of recent simulation results concerning the upsilon family of
$b\overline b$~mesons. This is an attractive system to study because so
much is known about it. The quark potential model, for example, provides an
accurate phenomenological model of the internal structure of the mesons.
Also there is  much experimental data for these mesons. The
low-lying states are largely insensitive to light-quark vacuum
polarization\,---\,for example, the $\Upsilon$, $\Upsilon^\prime$,
$\chi_b$\ldots{}are all far below the threshold for decays into
$B$~mesons\,---\,and therefore can be accurately simulated with $n_f=0$.
Furthermore, these mesons are very small, the~$\Upsilon$ being about five
times smaller than a light hadron. This makes them an excellent testing
ground for our ideas concerning large lattice spacings.

The $\Upsilon$~spectrum indicates that the $b$-quarks in the meson are
nonrelativistic. This means that the most important momentum
scales governing the meson's dynamics are smaller than the quark's
mass~$M_b$. We can take advantage of this fact by choosing an inverse lattice
spacing of order the quark mass, thereby excluding relativistic states from
the theory. Then it is efficient to analyze the heavy-quark dynamics using
a nonrelativistic lagrangian (NRQCD). The lagrangian used to generate the
results shown below was
\be
\lag_{\rm NRQCD} =
 \psi^\dagger \left(1\!-\!\frac{aH_0}{2n}\right)^{n}
 U^\dagger_{\hat t}
 \left(1\!-\!\frac{aH_0}{2n}\right)^{n}\left(1\!-\!a\delta H\right) \psi
 -\psi^\dagger\psi,
 \ee
where $n=2$, $H_0$ is the nonrelativistic kinetic-energy operator,
 \be
 H_0 = - {\delsq\over2\Mbz},
 \ee
$\Mbz$~is the bare quark mass, and $\delta H$ is the leading relativistic and
finite-lattice-spacing correction,
 \begin{eqnarray}
\delta H
&=& - \frac{(\delsq)^2}{8(\Mbz)^3}\left(1+\frac{a\Mbz}{2n}\right)
    + \frac{a^2\delfour}{24\Mbz} \nl
& & - \frac{g}{2\Mbz}\,\sigmav\cdot\Bv
            + \frac{ig}{8(\Mbz)^2}\left(\delv\cdot\Ev - \Ev\cdot\delv\right)
\nl & & - \frac{g}{8(\Mbz)^2} \sigmav\cdot(\delv\times\Ev - \Ev\times\delv) .
\label{deltaH}
\end{eqnarray}
Here $\delv$ and $\delsq$ are the simple gauge-covariant lattice derivative
and laplacian, while $\delfour$ is a lattice version of the continuum
operator $\sum D_i^4$. The chromoelectric and chromomagnetic fields,
$\Ev$~and~$\Bv$, are defined in terms of different components of
\be
\Uplaq\mu\nu - \Uplaq\mu\nu^\dagger.
\ee
The entire  action was tadpole
improved by dividing every link operator~$U_\mu$ by $u_0$.  Potential models
indicate that corrections beyond~$\delta H$ contribute only of order
5--10~MeV to $\Upsilon$ energies. The simulation results shown below used
gluon fields generated at~$\beta=6$, which corresponds to a lattice spacing
of about 1/12~fm or  about half the radius of an~$\Upsilon$.

The details of this lagrangian are  unimportant to us here. What matters is
that $\delta H$ consists of correction terms just like the ones we have
been analyzing for other theories, the only difference here being that we
are correcting both for finite-$a$ and for the absence of relativity (ie,
order~$v^2/c^2$ errors). The coefficients of the correction terms were
determined with tree-level perturbation theory and tadpole improvement,
using precisely the techniques outlined in the earlier sections. So the
extent to which~$\delta H$ improves the simulation results is a measure of
the efficacy of all the techniques discussed in these lectures.

\begin{figure}
\begin{center}
\setlength{\unitlength}{.02in}
\begin{picture}(80,140)(0,930)
\put(15,940){\line(0,1){120}}
\multiput(13,950)(0,50){3}{\line(1,0){4}}
\multiput(14,950)(0,10){10}{\line(1,0){2}}
\put(12,950){\makebox(0,0)[r]{9.5}}
\put(12,1000){\makebox(0,0)[r]{10.0}}
\put(12,1050){\makebox(0,0)[r]{10.5}}
\put(12,1060){\makebox(0,0)[r]{GeV}}

\put(40,940){\makebox(0,0)[t]{$\Upsilon$}}
\multiput(33,946)(3,0){5}{\line(1,0){2}}
\put(40,946){\circle*{2}}

\multiput(33,1002)(3,0){5}{\line(1,0){2}}
\put(40,1004){\circle*{2}}
\put(40,1004){\line(0,1){2}}
\put(40,1004){\line(0,-1){2}}

\multiput(33,1036)(3,0){5}{\line(1,0){2}}
\put(40,1032){\circle*{2}}
\put(40,1032){\line(0,1){7}}
\put(40,1032){\line(0,-1){7}}

\put(65,940){\makebox(0,0)[t]{$h_b$}}

\multiput(58,990)(3,0){5}{\line(1,0){2}}
\put(65,987){\circle*{2}}
\put(65,987){\line(0,1){3}}
\put(65,987){\line(0,-1){3}}

\multiput(58,1026)(3,0){5}{\line(1,0){2}}
\put(65,1025){\circle*{2}}
\put(65,1025){\line(0,1){9}}
\put(65,1025){\line(0,-1){9}}
\end{picture}
\end{center}
\caption{NRQCD simulation results for the spectrum of the
$\Upsilon (^3S_1)$ and $h_b (^1P_1)$ and their radial excitations. Experimental
values (dashed lines) are indicated for the $S$-states, and for the
spin-average of the $P$-states. The energy zero for the
simulation results is adjusted to give the correct mass to the $\Upsilon$.}
\label{spectups}
\end{figure}
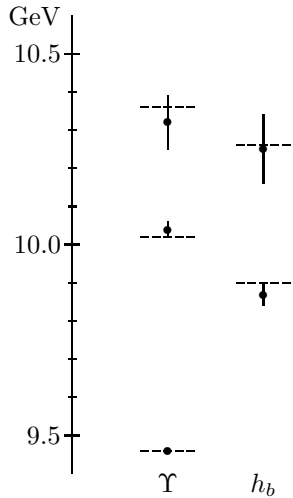

The $\Upsilon$~spectrum is very well described by the simulation. Simulation
results for the low-lying~${}^3S_1$ and~${}^1P_1$ energies are shown in
Figure~\ref{spectups}. These compare well with experimental results (the
horizontal lines), as they should since systematic errors are estimated to
be less than 20--40~MeV. It is important to realize that these are
calculations from first principles. The only inputs are the lagrangians
describing gluon and quark dynamics, and the only parameters are the bare
coupling constant and the bare quark mass. In particular, these results are
not based on a phenomenological quark-potential model. These are among the
most accurate lattice results to date.

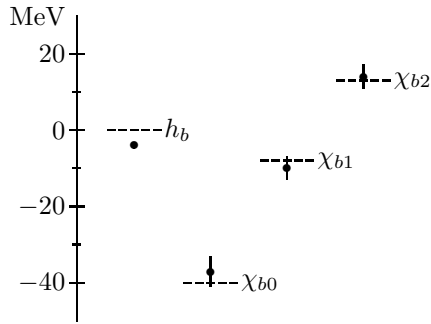
\begin{figure}
\begin{center}
\setlength{\unitlength}{0.02in}
\begin{picture}(110,90)(0,-50)
\put(15,-50){\line(0,1){80}}
\multiput(13,-40)(0,20){4}{\line(1,0){4}}
\multiput(14,-40)(0,10){7}{\line(1,0){2}}
\put(12,-40){\makebox(0,0)[r]{$-40$}}
\put(12,-20){\makebox(0,0)[r]{$-20$}}
\put(12,0){\makebox(0,0)[r]{$0$}}
\put(12,20){\makebox(0,0)[r]{$20$}}
\put(12,30){\makebox(0,0)[r]{MeV}}

\multiput(43,-40)(3,0){5}{\line(1,0){2}}
\put(58,-40){\makebox(0,0)[l]{$\chi_{b0}$}}
\put(50,-37){\circle*{2}}
\put(50,-37){\line(0,1){4}}
\put(50,-37){\line(0,-1){4}}

\multiput(23,0)(3,0){5}{\line(1,0){2}}
\put(38,0){\makebox(0,0)[l]{$h_b$}}
\put(30,-4){\circle*{2}}
\put(30,-4){\line(0,1){0.7}}
\put(30,-4){\line(0,-1){0.7}}

\multiput(63,-8)(3,0){5}{\line(1,0){2}}
\put(78,-8){\makebox(0,0)[l]{$\chi_{b1}$}}
\put(70,-10){\circle*{2}}
\put(70,-10){\line(0,1){3}}
\put(70,-10){\line(0,-1){3}}

\multiput(83,13)(3,0){5}{\line(1,0){2}}
\put(98,13){\makebox(0,0)[l]{$\chi_{b2}$}}
\put(90,14){\circle*{2}}
\put(90,14){\line(0,1){3}}
\put(90,14){\line(0,-1){3}}
\end{picture}
\end{center}
\caption{Simulation results for the spin structure of the lowest lying
$P$-state in the $\Upsilon$~family. The dashed lines are the experimental
values for the triplet states, and the experimental spin average of all states
for the singlet ($h_b$).}
\label{fsups}
\end{figure}

The corrections in~$\delta H$ have only a small effect on the overall
spectrum, but the spin structure is strongly affected. The lagrangian
without~$\delta H$ is spin independent, and gives no spin splittings at all.
Simulation results for the spin structure of the lowest lying $P$~state are
shown in Figure~\ref{fsups}. Again these compare very well with the data,
giving strong evidence that corrected lagrangians work. Systematic errors
here are estimated to be of order~5~MeV. Note that the spin terms in~$\delta
H$ all involve either chromoelectric or chromomagnetic field operators. These
operators are built from products of four link operators and so tadpole
improvement increases their magnitude by almost a factor of two
at~$\beta=6$. Without tadpole improvement of the spin splittings from this
simulation would have been much too small.

The correction terms also affect the simulation mass of
the~$\Upsilon$. Our nonrelativistic action gives only part of the total
energy or mass of the meson. The full mass is obtained by adding the masses
of the quarks to the meson's nonrelativistic energy~$E_{\rm NR}$:
\be \label{mass1}
M_\Upsilon = 2 \left( Z_m \Mbz - E_0 \right) + E_{\rm NR}
\ee
where~$Z_m$ and~$E_0$ are ultraviolet renormalizations that are computed
using perturbation theory, $\Mbz$ is a tunable parameter of the theory,
and $E_{\rm NR}$ is computed in the simulation. The bare quark mass for
the simulation discussed here was tuned
so that this formula gave~$M_\Upsilon = 9.5(1)$~GeV (the correct value
is~9.46~GeV). The $\Upsilon$~mass was also determined a second way in the
simulation by computing the nonrelativistic
energy of an~$\Upsilon$ as a function of its momentum~$p$, and fitting its
low-momentum behavior to the form
\be
E_\Upsilon(\pv) = E_{\rm NR} + \frac{\pv^2}{2\,M_{\rm kin}} + \cdots.
\ee
This determined the kinetic mass~$M_{\rm kin}$ of the meson. In a purely
nonrelativistic theory the kinetic mass equals the sum of the quark masses.
Only when relativistic corrections are included is this mass shifted to
include the binding energy, giving~$M_\Upsilon$. The simulation
without~$\delta H$ gave~$M_{\rm kin} = 8.2(1)$~GeV, which is quite different
from the upsilon mass determined using~\eq{mass1}. With~$\delta H$, the
simulation gave~$M_{\rm kin} = 9.5(1)$~GeV, which is in excellent agreement.
All of the spin-independent pieces of~$\delta H$ contribute to the shift
in~$M_{\rm kin}$; once again we have striking evidence that
corrected actions work.

This simulation has only two parameters: the bare coupling constant and the
bare quark mass. These were tuned to fit experimental data. From the bare
parameters we can compute the renormalized coupling and mass. This
simulation implies that the renormalized or ``pole mass'' of the
$b$-quark is
\be
M_b = 4.94(15)~{\rm GeV}.
\ee
The renormalized coupling that is obtained
corresponds to
\be
\alpha_\msb^{(5)}(M_Z) = 0.112(5),
\ee
which agrees
with results from high-energy phenomenology and  is about as accurate. This
last result is striking: it shows that the QCD of hadronic structure and the
QCD of high-energy quark and gluon jets are really the same theory.

\section{Conclusion}
The $\Upsilon$~simulations described in the last section show that
lattice QCD can produce accurate results even when the lattice
spacing is as large as half the radius of a meson. For light hadrons this
would correspond to $a
\approx 0.4$~fm  or  $\beta\approx5.4$ (for simulations with~$n_f=0$). Our
analysis shows that perturbation theory is still reliable at
such distances, and so perturbative improvement of the
lagrangians used in simulations probably still works. Given that
finite-volume errors become manageable for lattice sizes of order 2--3~fm, it
seems likely that reliable simulations of full QCD are possible on lattices
as small as~$6^4$. Simulations on such small lattices are literally a
thousand times faster than simulations on the $20^4$~lattices commonly used
today. If coarse lattices really do work, the shift to small lattices and
improved lagrangians will have a revolutionary effect on numerical QCD.

\section{Acknowledgements}

Many of the ideas and opinions presented here grow out of my
long-standing collaboration with Paul Mackenzie. The numerical results
in Sections~3 and~4 are from our work on lattice perturbation theory.
The~$\Upsilon$ simulations in Section~5 were by myself and my
collaborators in the NRQCD~collaboration.
This work is supported by a grant
from the National Science Foundation.

\section{Bibliography}
The central point in these lectures is that errors due to finite lattice
spacing should be removed by improving the lagrangian, and not by
decreasing the lattice spacing. Improvement schemes have a long history;
see, for example,
 \begin{refer} K.\ Symanzik, {\em Nucl.\ Phys.\/} {\bf B226}, 187(1983);
\end{refer}
 \begin{refer} M.\ L\"uscher and P.\ Weisz, {\em Comm.\ Math.\ Phys.\/}
 {\bf 97}, 59(1985).\end{refer}
The implementation of these ideas has been
held back by persistent worries about the validity and utility of
perturbation theory; the improvement program is much more complicated
without perturbation theory.  Renormalized perturbation theory (Section~3)
and tadpole improvement (Section~4) appear to resolve these problems. These
techniques are developed at greater length in
\begin{refer} G.P. Lepage and P.B. Mackenzie, {\em Phys. Rev.\/} {\bf D48},
2250 (1993). \end{refer}

My own interest in improved lagrangians and renormalization group
strategies came from studying very different problems. Bill Caswell and I
discovered that renormalization techniques provide a powerful tool for
high precision analyses of QED boundstates like positronium. This led to
the development of nonrelativistic versions of QED and eventually also
of QCD. These developments are discussed in
\begin{refer} W.E.\ Caswell and G.P.~Lepage, {\em Phys.\ Lett.\/} {\bf 167B},
437 (1986). \end{refer}
\begin{refer} G.P.\ Lepage and B.A.\ Thacker, {\em Nucl.\ Phys.\/} {\bf
B(Proc. Suppl.)4}, 199(1988), and {\em  Phys.\ Rev.\/} {\bf D43},
196(1991).\end{refer}
\begin{refer} G.P. Lepage, L. Magnea, C. Nakhleh, U. Magnea, and K.
Hornbostel, {\em Phys.\ Rev.\/} {\bf D46}, 4052(1992).\end{refer}
\begin{refer} G.P. Lepage, "What is Renormalization?" in {\em From Actions
to Answers}, edited by T. DeGrand and D. Toussaint (World
Scientific, Singapore, 1989). \end{refer}
The last of these papers has a
pedagogical discussion of the ideas from renormalization theory that are
used in the present lectures. The NRQCD results presented in Section~6
are described in a series of papers to be published in the near
future (by C.~Davies et al).

For an elementary introduction to lattice QCD see
\begin{refer} T.-P. Cheng and L.-F. Li, {\em Gauge theory of elementary
particle physics\/} (Clarendon Press, Oxford, 1991). \end{refer}
At a more advanced level is
\begin{refer} M. Creutz, {\em Quarks, gluons and lattices\/} (Cambridge
University Press, Cambridge, 1985). \end{refer}
The generic behavior of lattice algorithms is reviewed in
\begin{refer} G.P. Lepage, "The Analysis of Algorithms for Lattice Field
Theory" in {\em From Actions to Answers}, edited by T. DeGrand and D.
Toussaint (World Scientific, Singapore, 1989). \end{refer}
Useful references concerning the Feynman rules for weak-coupling perturbation
theory in lattice QCD and some standard results are
\begin{refer} H. Kawai, R. Nakayama and K. Seo, {\em Nucl.\
Phys.\/} {\bf B189}, 40 (1981).\end{refer}
\begin{refer} U. Heller and F. Karsch, {\em Nucl. Phys.\/} {\bf B251}, 254
(1985). \end{refer}
Two good reviews of the current status of QCD simulations are:
\begin{refer} A. Ukawa, {\em Nucl.\ Phys.\/} {\bf B(Proc. Suppl.)30},
3(1993).  \end{refer}
\begin{refer} P.B. Mackenzie, in the {\em Proceedings of the 1993
Lepton-Photon Symposium}, edited by P. Drell and D. Rubin (AIP Press, New
York, 1994).\end{refer}
The first of these is from one of the annual meetings on lattice field
theory. The proceedings from these meetings are  primary references for
recent developments in the field.

\end{document}